\documentstyle[mnextra,psfig,graphicx]{mn}
\nobrackets
\def\mnras{MNRAS}

\def\apj{ApJ}
\def\apjs{ApJS}
\def\aap{A\&A}
\def\ARAA{ARA\&A}

\def\hMpc{\,h^{-1}\,\hbox{Mpc}}

\def\Msol{\hbox{M}_{\odot}}

\def\hMsol{h^{-1}\,\hbox{M}_{\odot}}
\def\erg{\,\hbox{erg}}

\def\icm2{\,\hbox{cm}^{-2}}
\def\icm3{\,\hbox{cm}^{-3}}
\def\gcm3{\,\hbox{g}\,\hbox{cm}^{-3}}

\def\kms{\,\hbox{km}\,\hbox{s}^{-1}}
\def\hkpc{\,h^{-1}\,\hbox{kpc}}

\def\kmsMpc{\,\hbox{km}\,\hbox{s}^{-1}\,\hbox{Mpc}^{-1}}

\def\h50{\, h_{50}}

\def\sl0{\epsilon_{0}}

\def\H0{H_0=100 \, h \, {\rm kms^{-1}Mpc^{-1}}}
\def\Zsol{Z_{\odot}}
\def\ln{{\rm ln}}
\def\log{{\rm log}}
\def\NSPH{N_{\rm SPH}}

\def\kelvin{\hbox{K}\,}
\def\Omegab{\Omega_{\rm b}}

\def\AP3M{$\rm{AP}^3\rm{M}$\,}
\def\p3m{{\rm P}$^{3}${\rm M}\,}
\def\ap3m{{\rm AP}$^{3}${\rm M}\,}
\def\hydra{{\sc hydra} }

\def\rhocr0{\rho_{\rm cr,0}}

\def\Myr{\hbox{Myr}}

\def\etal{{et al.\thinspace}}
\def\eg{{e.g.\thinspace}}

\def\ie{{i.e.\thinspace}}
\def\cf{{c.f.\thinspace}}

\begin{document}
\journal{Submitted to MNRAS; preprint astro-ph/0106462}

\title [Star formation and supernova feedback in cosmological simulations]
       {Including star formation and supernova feedback within 
         cosmological simulations of galaxy formation}

\author[Scott T. Kay et al.]
        {Scott T. Kay,$^{1,2}$\thanks{email: S.T.Kay@sussex.ac.uk}
          Frazer R. Pearce,$^{1,3}$ Carlos S. Frenk$^{1}$ 
          and Adrian Jenkins$^{1}$\\
        $^{1}$ Physics Department, University of Durham, Science Laboratories,
        South Road, Durham DH1 3LE\\
        $^{2}$ Astronomy Centre, CPES, University of Sussex, Falmer, Brighton 
        BN1 9QJ\\
        $^{3}$ Physics and Astronomy, University of Nottingham, Nottingham 
        NG7 2RD\\}
      
\date{\today}
\maketitle

\begin{abstract}

We investigate phenomenological models of star formation and supernova
feedback in $N$-body/SPH simulations of galaxy formation. First, we
compare different prescriptions in the literature for turning cold gas
into stars neglecting feedback effects. We find that most
prescriptions give broadly similar results: the ratio of cold gas to
stars in the final galaxies is primarily controlled by the range of gas
densities where star formation is allowed to proceed efficiently. In
the absence of feedback, the fraction of gas that cools is much too
high resulting, for example, in a $K$--band luminosity function that
is much brighter than observed. This problem is ameliorated by
including a feedback model which either imparts radial kinetic
perturbations to galactic gas or directly reheats such material and
prevents it from cooling for a certain period of time. In both these
models, a significant fraction of cold gas is heated and expelled from
haloes with an efficiency that varies inversely with with halo circular
velocity. Increasing the resolution of a simulation allows a wider
dynamic range in mass to be followed, but the average properties of
the resolved galaxy population remain largely unaffected.  However, as
the resolution is increased, more and more gas is reheated by small
galaxies; our results suggest that convergence requires the full mass
range of galaxies to be resolved.
\end{abstract}


\section{Introduction}
\label{sec:intro}

The first detailed modelling of galaxy formation more than twenty
years ago (\cite{WR}) revealed a significant difficulty in
hierarchical clustering cosmologies. These assume that galaxies form
when gas cools and condenses in dark matter haloes. However, at early
times, pre-galactic gas clouds are so dense that the cooling time is
extremely short. Consequently, in the absence of other processes,
baryonic matter cools catastrophically in sub-galactic haloes and
forms stars, leaving no gas available to make up the intergalactic
medium observed at high redshift, or the intracluster medium observed
at low redshift. White \& Rees proposed a solution to this problem:
energy released from stars in the course of their evolution would act
as negative feedback on the gas, limiting its cooling rate and
associated star formation.

These ideas carry through immediately to modern cosmological theory, 
which is based upon the cold dark matter paradigm (\cite{WF}). 
This provides a natural motivation for hierarchical clustering and 
so it inherits the ``cooling catastrophe'', generic to this kind of 
models. There is now a growing body of work attempting to understand 
likely sources of feedback (\eg \cite{Dekel&Silk86}; \cite{Efstathiou1992};
\cite{Thoul&Weinberg96}; \cite{Silk&Rees98}; \cite{Maclow&Ferrara}; 
\cite{Efstathiou2000}; \cite{Madau2000}), their effects on models of galaxy 
formation (\eg \cite{WF}; \cite{Cole91}; \cite{K92}; \cite{NW93}; \cite{GK94}; 
\cite{Cole94}; \cite{KWH96}; \cite{YKKK}; \cite{SP98};
\cite{Thacker&Couchman2000}; 
\cite{Cole2000}), and on the chemical enrichment of the intergalactic medium
at high redshift (\eg \cite{Ellison99}; \cite{Schaye2000}). Efforts to
detect observational signatures of feedback are also being made (\eg
\cite{Martin1999}; \cite{Theuns2000}; \cite{Pettini2001}).

Recent modelling of galaxy formation, including the effects of
feedback, has made use of two related techniques, semi-analytic
modelling and direct numerical simulation. The first is a direct
descendant of the methodology developed by White \& Rees (\shortcite{WR})
and provides a computationally efficient way to calculate a detailed
model of galaxy formation. The merger history of dark matter haloes is
followed using either statistical techniques (e.g. \cite{GK94}; Cole
\etal \shortcite{Cole94}, \shortcite{Cole2000}; \cite{SP98}) or $N$-body simulations
(e.g. \cite{kns}; \cite{Benson01};
\cite{somerville00}), the evolution of gas is treated 
by means of a spherically symmetric cooling flow model, star
formation and the effects of feedback are parameterised according to
simple, observationally motivated rules and stellar evolutionary
effects are incorporated by means of a stellar population synthesis
model. Chemical evolution and dust extinction or emission are readily
incorporated in this scheme (e.g. \cite{Granato00}). Semi-analytic
modelling has proved to be extremely successful for the interpretation
of observational data at both low and high redshift, as reviewed
recently by Baugh \etal (\shortcite{Baugh01}).

In this paper, we are concerned with the treatment of star formation and
feedback in gasdynamical simulations of galaxy formation.  Numerical
simulations have the advantage that some of the assumptions made in the
semi--analytic treatment of gas dynamics can be relaxed. In particular, no
artificial symmetries are imposed on a situation which is inherently
asymmetric. However, the application of simulation techniques to galaxy
formation is still in its infancy because it places strenuous demands on
computational resources, consequently limiting the dynamic range of
resolved structures. Early work, while making brave first attempts at
simulating cosmological galaxy formation, was hampered to varying degrees
by relatively poor resolution (\eg
\cite{CO92}; \cite{KHW}; 
\cite{ESD}; \cite{KWH96}; \cite{FEWS}).  It is only over the last few
years that parallel supercomputers have enabled simulations large
enough to resolve present-day bright galaxies adequately within
statistically-relevant volumes (\eg \cite{Pearce99};
\cite{Fardal2000}; \cite{Pearce2000}). Detailed tests of varying
assumptions and parameter values in simulations based on the Smooth
Particle Hydrodynamics technique are presented in Kay \etal 
(\shortcite{Kay2000}, hereafter K2000)

It is unfeasible to consider an explicit treatment of star formation
and feedback within a cosmological simulation because these processes
occur on much smaller scales than can be resolved with present
techniques. The only approach (which is currently also used in
semi--analytic models) is to develop phenomenological {\it
prescriptions} and test their validity by comparing their effects to
as many observables as possible. The main aim of this paper is to
compare the outcome of adopting various star formation and feedback
prescriptions in a series of small {\it test} simulations, similar to
those analysed by K2000.  We assume that feedback is produced only as
a result of energy being injected into gas through supernovae occurring 
in the star forming regions that develop in the simulations. We also
investigate the effects of varying the numerical resolution, albeit
over a small dynamic range. Our primary concern is the effect of
feedback on the properties of the resolved galaxy population, mainly
at $z=0$, but we also examine the properties of reheated gas, since it
is these predictions that may provide some of the most stringent
observational tests.

We begin with a short description of our numerical methods in
Section~\ref{sec:numerical_method}. We compare various star formation
models in Section~\ref{sec:sf_starform} before investigating feedback
in Section~\ref{sec:sf_feedback}, focusing on both the galaxies and
the distribution of reheated gas. The sensitivity of our results to
numerical resolution is explored in Section~\ref{sec:sf_resol}.
Finally, we draw conclusions in Section~\ref{sec:sf_conclude}.

\section{Numerical method}
\label{sec:numerical_method}

\subsection{The code}
\label{subsec:code}

The simulations were carried out using the \hydra code
\footnote{\hydra is publicly available and can be downloaded from 
{\tt http://hydra.mcmaster.ca}}(\cite{CTP}). This 
combines an adaptive particle-particle/particle-mesh (\ap3m) algorithm
to calculate gravitational forces (\cite{COUCH}) with Smoothed
Particle Hydrodynamics (SPH; \eg \cite{BENZ}; \cite{MON} and
references therein) to estimate hydrodynamical quantities. The SPH
implementation differs from that described in Couchman, Thomas \&
Pearce (\shortcite{CTP}), as discussed by Thacker \& Couchman
(\shortcite{Thacker&Couchman2000}).  In addition to these changes, the
SPH neighbour search algorithm in \hydra has been modified in order to
avoid setting an artificial lower limit on the gas density, present
in older versions of the code (which used a single chaining mesh to find
neighbours for both the PP gravity and SPH calculations). 
In the new version, a separate neighbour list for the gas particles is 
constructed using the algorithm described in Appendix~A2 of Theuns \etal 
(\shortcite{Theuns98}).

Radiative cooling is modelled in the manner described by Thomas
\& Couchman (\shortcite{THOMCOUCH}), except that tabulated cooling rates 
(as a function of temperature and metallicity) from Sutherland \& Dopita
(\shortcite{SUTDOP}) are implemented. This assumes that the gas is in 
collisional ionization equilibrium and is applicable only to radiative 
processes at temperatures above $10^4\kelvin$. Contributions to the 
cooling rate at temperatures below $10^4\kelvin$ are ignored, as are the 
radiative heating contributions from a photo-ionizing background. The 
former are only important in structures on scales significantly below the
resolution of our simulations. Although photo-heating is
important in determining the structure of the low density gas
responsible for the Ly-$\alpha$ forest, previous studies have shown
that the process is unimportant for galaxies forming in haloes with circular 
velocities above $\sim 50 \kms$ (\eg \cite{QKE}; \cite{NS97}; \cite{WHK97}), 
as is the case in our simulations. 

\subsection{Cosmological models}
\label{subsec:cosmology}

\begin{table}
\caption{Values of key parameters of the cosmological models studied
in this paper} 
\begin{center}
\begin{tabular}{cccccccc}
Model & $\Omega_0$ & $\Lambda_0$ & $\Omegab$ & $h$ & $\Gamma$ &
$\sigma_8$ & $z_i$\\
\hline
SCDM & 1 & 0 & 0.06 & 0.5 & 0.5 & 0.6 & 24\\
$\Lambda$CDM & 0.35 & 0.65 & 0.038 & 0.71 & 0.21 & 0.9 & 49\\
\end{tabular}
\label{tab:sf_back_param}
\end{center}
\end{table}

We carried out simulations using two cold dark matter models: SCDM and
$\Lambda$CDM. The former assumes a critical density universe, 
$\Omega_0=1$ and $\Lambda_0=0$, Hubble constant, in units of $100
\kmsMpc$, $h=0.5$ and baryon density, 
$\Omega_b=0.06$. The $\Lambda$CDM model assumes a flat, low density
universe, with $\Omega_0=0.35$ and $\Lambda_0=0.65$, Hubble constant,
$h=0.71$, and baryon density, $\Omega_b=0.038$.  Current determinations of
cosmological parameters (from, for example, microwave background
anisotropies, high redshift supernovae, and the baryon fraction in galaxy
clusters) favour the $\Lambda$CDM model.  However, we do not expect our
conclusions to differ significantly in other hierarchical models. 

Initial density fluctuations were calculated for the SCDM model using
the approximate transfer function given by Bond \& Efstathiou
(\shortcite{BE}) with shape parameter, $\Gamma=
\Omega_0h=0.5$ and amplitude set by $\sigma_8=0.6$ (as required for
agreement with the present day abundance of rich clusters; \eg
\cite{VL}; \cite{ECF}). For the $\Lambda$CDM model, the transfer
function of Bardeen \etal (\shortcite{BBKS}) was adopted and $\Gamma$
was set to 
$\Gamma=\Omega_0 h \, \exp (-\Omega_b/\Omega_0(\Omega_0+\sqrt{2h}))=0.21$ 
(\cite{Sugi95}).  The amplitude of density
fluctuations was normalised by setting $\sigma_8=0.9$, as required by the
cluster abundance argument in this case. 

The gas particle mass was held fixed for all simulations within each
cosmological model, at $\sim 5 \times 10^8 \hMsol$ in the case of SCDM and
to $\sim 3 \times 10^8 \hMsol$ in the case of $\Lambda$CDM. Both models
assumed an initial (equivalent Plummer) softening length of $\sim 20 \hkpc$
which was fixed in comoving coordinates until $z\sim 1$, after which it
was fixed in physical coordinates to $10 \hkpc$ until $z=0$.  The SCDM
simulations were started at a redshift of 24 and the $\Lambda$CDM
simulations at a redshift of 49 and both sets were run to $z=0$.

\section{Star formation}
\label{sec:sf_starform}

\begin{table}
\begin{center}
\caption{Summary of the star formation methods studied}
\begin{tabular}{ll}  
Simulation & Details \\
\hline
{\sc pearce1}    & $\delta>5000$  ; \ $T<10^{5}\kelvin$\\
{\sc pearce2}    & $\delta>50,000$ ;\ $T<10^{5}\kelvin$\\
{\sc summers1}   & $\delta>10$ ;\ $\rho>10^{-25}\gcm3$ ;\  $T<3\times10^{4}\kelvin$\\ 
{\sc summers2}   & $\delta>10$ ;\ $\rho>10^{-26}\gcm3$ ;\ $T<3\times10^{4}\kelvin$\\
{\sc navarro}   &  $\nabla.{\bf v}<0 ;\ \rho>7\times10^{-26}\gcm3$\\
{\sc katz}      &  $\nabla.{\bf v}<0 ;\ \delta>10 ;\ \rho>10^{-25}\gcm3 ;\ 
\tau_{\rm J}>\tau_{\rm dyn}$\\ 
{\sc mihos}      & $\nabla.{\bf v}<0 ;\ \delta>10$ ;\ 
$C_{\rm sfr}=5\times10^{-4}$ ;\ $N=3/2$\\
\end{tabular}
\label{tab:sfdetails}
\end{center}
\end{table}

In this section we concentrate on comparing several star formation models
drawn from the literature.  We present results for SCDM simulations with
$32768$ gas and $32768$ dark matter particles, initially perturbed from a
regular grid, within a $10 \hMpc$ comoving box. The initial conditions and
parameters are identical to those of the fiducial simulation of K2000,
where further details may be found. Although the box-size is too small to
allow statistically robust results for the galaxy population, the
simulations are only used for comparative purposes.

\subsection{Star formation models}
\label{subsec:sf_star_models}
We now summarise each star formation model studied.  In all cases, the
basic procedure is the same: at the end of each timestep, a subset of gas
particles is identified as being eligible to form stars.  A fraction of
them are then converted into stars, which are subsequently subject only to
gravitational interactions.  Details of each method are given in
Table~\ref{tab:sfdetails}.

\subsubsection{The Pearce method}
\label{subsubsec:pearce}

This is based on the method used by Pearce (\shortcite{PSF}). Potential
star particles are identified with gas particles having overdensity,
$\delta>550$ and temperature, $T<10^{5}\kelvin$. We carried out two
simulations, both using the same temperature threshold as Pearce but one
with an overdensity threshold of 5000 (labelled {\sc pearce1}) and the
other with a threshold of 50,000 (labelled {\sc pearce2}).  Our chosen
overdensity thresholds are higher than that used by Pearce, who selected a
value in order to ensure that a certain fraction of gas was converted into
stars by $z=0$. However, Pearce's value was also close to the maximum
resolved SPH gas overdensity in his simulations.  This limit is estimated
by considering a sphere of radius $2h_{\rm min}$ containing $N_{\rm SPH}$
neighbours. (The minimum SPH smoothing length, $h_{\rm min}$ is set equal to
1.17 times the equivalent Plummer softening length.) For our choice of
parameters ($2h_{\rm min}=0.0234
\hMpc$ and $N_{\rm SPH}=32$), this limit is $\delta \sim 20,000$ at $z=0$.
We deliberately set our choice of overdensity thresholds to straddle this
resolution limit.

\subsubsection{The Summers method}

In the method used by Summers (\shortcite{Summers_phd}) gas particles
are converted into stars if they satisfy a temperature criterion,
$T<3\times 10^4\kelvin$ and a {\it physical} density criterion,
$n_{\rm H} > 0.1 \icm3$. Additionally, eligible gas particles must be
within overdense regions, $\delta > 10$, so that star formation does
not occur within shallow potential wells at high redshift. We
performed two simulations using the Summers method, labelled {\sc
summers1} and {\sc summers2} respectively. For {\sc summers1}, we used
a density criterion, $\rho >10^{-25} \gcm3$ ($\rho=m_{\rm H} \, n_{\rm
H}/X$; we assume $X=0.76$), approximately a factor of 2 lower than
used by Summers. This threshold corresponds to a baryon overdensity,
$\delta \sim 3.5 \times 10^5$ at $z=0$ which is over an order of
magnitude higher than our resolution limit. Because of this, we
carried out a second simulation ({\sc summers2}) for which we chose a
density threshold of $\rho>10^{-26}\gcm3$.

\subsubsection{The Navarro \& White method}

We have also studied the method of Navarro \& White (\shortcite{NW93})
who applied it to follow star formation in disk galaxies. This is a
popular method that has since been used by several other authors, with
minor modifications (\eg \cite{SM95}; \cite{TISSERA};
\cite{Carraro98}; \cite{BUONOMO99}). To summarise, regions of star
formation are identified in which the gas is in a convergent flow 
(${\bf \nabla.v} < 0$) and has a density, $\rho > 7 \times 10^{-26} \gcm3$. 
Gas particles satisfying these criteria are split into equal mass gas and 
star particles, after a time interval, $\Delta t_*= \ln 2 \, \tau_{\rm dyn}$, 
where  $\tau_{\rm dyn} \propto 1/\sqrt{\rho}$ is the local dynamical
time of the gas. When the gas mass is less than 5 per cent of its original 
value, the remaining fraction is converted into stars. We label the
simulation using this method {\sc navarro}.

\subsubsection{The Katz, Weinberg \& Hernquist method}

The fourth method investigated is that used in the cosmological
simulations of Katz, Weinberg \& Hernquist (\shortcite{KWH96}; see
also \cite{K92}). Four criteria are considered to decide if a gas
particle is eligible to form stars: the divergence criterion, as used
in the Navarro \& White method; the physical density threshold used as
in the Summers method; an overdensity threshold, $\delta>55.7$, and a
Jeans instability condition, $\tau_{\rm J} > \tau_{\rm dyn}$, where
$\tau_{\rm J}\equiv h/c_s$ is a ``Jeans'' timescale, $h$ is the
particle's SPH smoothing length and $c_s=(\gamma(\gamma-1)u)^{1/2}$
\footnote {$u=kT/(\gamma-1) \,\mu m_{\rm H}$; we take 
the ratio of specific heats at constant volume and constant pressure
for a monatomic ideal gas to be $\gamma=5/3$, and the mean molecular
weight of a gas of primordial composition to be $\mu=0.59$} is the
adiabatic sound-speed of the gas. (Note that this condition depends on
the resolution of the simulation since $h$ is limited by the
gravitational softening.) If a gas particle is eligible to form stars,
a third of its mass is converted if $r<1-\exp(-\Delta t/{\rm
min}(\tau_{\rm dyn}, \tau_{\rm cool}))$, where $r$ is a random number
between 0 and 1, $\Delta t$ is the timestep and $\tau_{\rm cool}
\propto 1/\rho$ is the cooling time. This method gives each particle a
dual identity (we refer to it as a {\it schizophrenic} particle),
allowing it to carry both a gas mass and a stellar mass.
Gravitational forces are calculated using the total mass and
hydrodynamical forces using only the gas mass. Once the gas mass has
dropped below $5$ per cent of its original value, the particle becomes
a purely collisionless star particle. We refer to the simulation using
this method as {\sc katz}.

\subsubsection{The Mihos \& Hernquist method}
\label{subsubsec:mihos}

The final model studied is based on the method used by Mihos \&
Hernquist (\shortcite{Mihos&Hernquist}). This uses the Schmidt law
(\cite{schmidt}; see also \cite{KEN98} and references therein), an
empirical (power-law) relation between the star formation rate per
unit area and the surface density of H{\sc i} gas in late-type
galaxies. Mihos \& Hernquist convert this to a 3-dimensional relation,
such that the fractional change in mass converted into stars during a
timestep is $\Delta M_*/M_{\rm gas} = C_{\rm sfr} \, \rho^{N-1} \,
\Delta t$.  This equation is applied to gas particles with $\delta>10$
and ${\bf \nabla.v}<0$, although rather than converting part of a
particle's mass into stars, we convert all of it if $\Delta
M_*/M_{\rm gas}>r$, where $r$ is a random number between 0 and 1. We
study a simulation using this method, labelled {\sc mihos}, in which
we have set $N=3/2$, consistent with the observed value, and $C_{\rm
sfr}=5\times 10^{-4} \,{\rm cm}^{3/2}{\rm g}^{-1/2}{\rm s}^{-1}$.

\subsection{Results}
\label{subsec:sf_stars_results}

\subsubsection{General properties at $z=0$}
\label{subsubsec:sf_stars_gen}

\begin{table}
\begin{center}
\caption{General properties of the simulations with star formation at $z=0$}
\begin{tabular}{lccccc}  
Simulation & CPU & $f_{\rm cool}$ & $f_*$ & $N_{\rm gal}$\\
\hline
{\sc gas}       & 1.00  & 0.35  & 0.00  & 74\\
{\sc pearce1}   & 0.32  & 0.33  & 0.28  & 64\\
{\sc pearce2}   & 0.51  & 0.34  & 0.25  & 70\\
{\sc summers1}  & 0.56  & 0.34  & 0.16  & 72\\ 
{\sc summers2}  & 0.37  & 0.34  & 0.26  & 70\\
{\sc navarro}   & 1.33  & 0.35  & 0.19  & 69\\
{\sc katz}      & 0.67  & 0.35  & 0.15  & 71\\
{\sc mihos}     & 0.48  & 0.35  & 0.24  & 64\\
\end{tabular}
\label{tab:sfresults}
\end{center}
\end{table}

Table~\ref{tab:sfresults} lists general properties of the 7
simulations at $z=0$, and of the equivalent simulation without star
formation (labelled {\sc gas}). Column~2 gives the CPU time required
to complete each run, relative to {\sc gas}.  (To provide
a fair comparison, differences in the machine-dependent clock rate
were removed by dividing the total time taken for each simulation by
the average time per step over the first 10 steps.)  The least
efficient method is {\sc navarro}, which takes around a third longer
to complete than {\sc gas} because of the increased number of
particles produced by star formation. The rest of the simulations all
take around half of the time to complete than {\sc gas}, with the
shortest being {\sc pearce1}. Including a prescription for star
formation reduces the CPU requirement because of the reduced SPH
workload.

Column 3 gives the fraction of baryons at $z=0$ in stars or gas with
$\rho>500\left<\rho\right>$ and $T<30,000 \kelvin$, labelled $f_{\rm
cool}$.  Gas that satisfies these criteria (\ie it is cold and dense)
is readily associated with galaxies, since the two thresholds
enclose a region of the $\rho-T$ plane containing material which has
radiatively cooled (see, for example, \S3.2 of K2000).  For all
simulations, including {\sc gas}, $f_{\rm g}=0.34 \pm 0.01$. As found
by Pearce (\shortcite{PSF}), the conversion of cold dense gas into
stars does not significantly affect the amount of gas that cools.

Column 4 gives the total fraction of baryons in stars, $f_*$, for each
run. The range of $f_*$ is much wider than for $f_{\rm cool}$, varying from
15 per cent ({\sc katz}) to 28 per cent ({\sc pearce1}). Increasing the
overdensity threshold from 5000 ( {\sc pearce1}) to 50,000 ({\sc pearce2})
decreased $f_*$ by only 0.03. This small difference is due to the extra
time it takes for cooled material to collapse to a higher overdensity
before forming stars. {\sc Summers2} produced a similar fraction of stars
as {\sc pearce1} and {\sc pearce2}, but 40\% more than {\sc summers1}. The
difference in the {\sc summers} simulations is also due to the choice of
density threshold which, at $z=0$, is about an order of magnitude higher
for {\sc summers1} than for {\sc pearce2}. {\sc katz} and {\sc navarro}
produce similar values of $f_*$ (to within 20\%) to {\sc summers1} because 
they use the same density threshold. The number of stars forming at any
given time in {\sc mihos} is controlled by the choice of normalisation
constant, $C_{\rm sfr}$. The value, $C_{\rm sfr}=5\times10^{-4} \,{\rm
cm}^{3/2}{\rm g}^{-1/2}{\rm s}^{-1}$, was chosen (by trial and error) to
give a similar fraction of stars at $z=0$ as {\sc pearce2}.

The final column in Table~\ref{tab:sfresults} lists the number of galaxies
in each simulation. These were identified using a friends-of-friends
group-finder (\cite{DEFW}) on the cold gas and star particles (\ie on those
used to calculate $f_{\rm cool}$ above), with a linking length, $b=0.1$
(comparable to the gravitational softening length.) Any objects with fewer
star particles than $\NSPH$ (32 for the simulations studied in this paper)
were discarded.  All runs contain $\sim 70$ galaxies at $z=0$ to within
10\%.
%There are no obvious systematic trends in the number of galaxies
%with other general properties.

\subsubsection{Star formation histories}
\label{subsubsec:sfr}

\begin{figure}
\centering
\centerline{\psfig{file=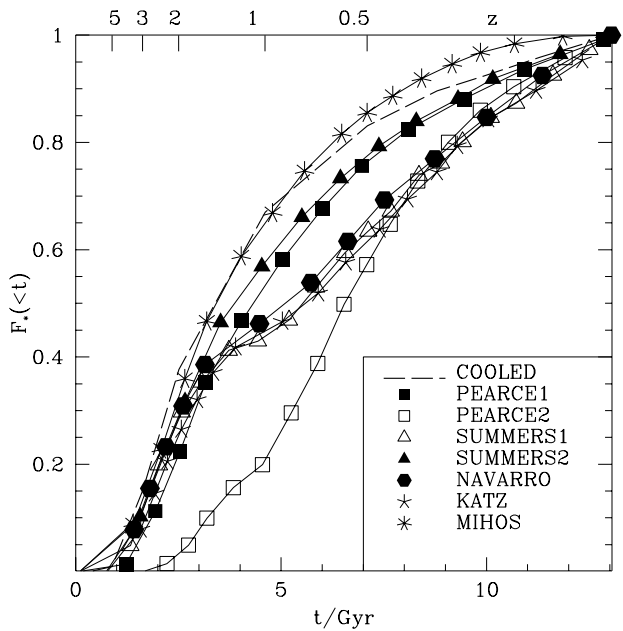,height=8.5cm}}
\caption{Cumulative fraction of baryons in stars, normalised to the 
$z=0$ value, plotted against time. Also shown is the 
normalised fraction of cooled material (defined in the text)
for the simulation without star formation. Redshift intervals are 
marked along the top of the figure.}
\label{fig:sfr}
\end{figure}

Fig.~\ref{fig:sfr} illustrates the star formation histories in the
simulations by means of the cumulative fraction of baryons turned into
stars as function of time, relative to the $z=0$ fraction (the corresponding 
redshift intervals are marked along the top of the figure). The cumulative 
fraction of cooled gas ($\rho>500\left<\rho\right>$ and $T<30,000 \kelvin$)
for the simulation without star formation, is also plotted.
The instantaneous star formation rate is proportional to the gradient of each
star fraction curve.

All simulations started cooling material and forming stars within the
first 2 Gyr ($z > 2$). The exact time when the first stars
form depends upon two factors: the resolution of the simulation, which
determines the formation time of the first resolved dark matter haloes,
and the range of gas densities where star formation occurs. Since the
resolution is fixed in all our simulations, only the second factor is
relevant. 

The star formation history in {\sc pearce1} is similar to that of cooled 
material, \ie it rises sharply from $z=5$ until $z \sim 2$, before
rolling over, implying a decreasing star formation rate at lower redshift.  
However, in {\sc pearce2}, the delay caused by a higher overdensity threshold 
causes star formation to begin later and to proceed at a slower rate until 
around 5 Gyr ($z \sim 1$), after which the $z=0$ normalised star
formation rate  is higher than in {\sc pearce1}. 

{\sc summers1}, {\sc navarro} and {\sc katz} have the same histories
at all times, demonstrating that the additional features of the latter
two models are irrelevant in determining the average star formation
rate within the simulations. (We exclude the latter two simulations
from further discussion since they produce nearly identical results
to {\sc summers} for all quantities studied in this paper.)

These curves are very similar to those
from the {\sc pearce} simulations until $z=1-2$, where a ``kink''
develops, due to the behaviour of the gravitational softening. At
early times, the softening is constant in comoving coordinates, and so
the minimum resolved length scale in physical coordinates grows in
proportion to the expansion factor, $a = 1/(1+z)$, leading to a {\it
decrease} (as $(1+z)^3$) in the maximum resolved physical
density. %(\cf Section~\ref{subsubsec:pearce}). 
By $z \sim 1.7$, this
density has dropped to $\rho \sim 10^{-26} \gcm3$, the {\sc summers1}
density threshold. This suppresses further star formation until $z=1$,
when the softening becomes constant in physical coordinates, allowing
the star formation rate to increase again. {\sc summers2} exhibits a
similar star formation history to {\sc pearce1}. In these cases no
kink is evident since the density thresholds are always lower than the
maximum resolved density.

Finally, in {\sc mihos}, the relative fraction of stars rises more
sharply than other curves at high redshift but flattens off by $z=0$. 
For this run, an approximate minimum density threshold can be calculated 
by first equating the relevant star formation timescale,
$\tau_{\rm sfr}=M_{\rm gas} \Delta t/\Delta M_* = 
1/C_{\rm sfr} \rho_{\rm gas}^{1/2}$ 
(\cf \S\ref{subsubsec:mihos}) to the lookback time for an $\Omega=1$
universe,  
$\tau_{\rm look} = (2/3H_0)(1-(1+z)^{-3/2})$; at higher densities, 
$\tau_{\rm sfr}<\tau_{\rm look}$.
Assuming that $\delta \sim \rho/\left<\rho\right>$ (\ie for $\delta>>1$),
the minimum overdensity scales as $\delta_{\rm min}
\propto ((1+z)^{3/2}-1)^{-2}$, a decreasing function of redshift. At high
redshift, this overdensity threshold was sufficiently low to allow
efficient star formation to take place. However, at lower redshift,
the decrease in the density of the Universe and in the available time
left to form stars before $z=0$, causes the number of gas particles
above this threshold to drop rapidly.

\subsubsection{Cold gas fractions}
\label{subsubsec:gal_cgf}

\begin{figure}
\centering
\centerline{\psfig{file=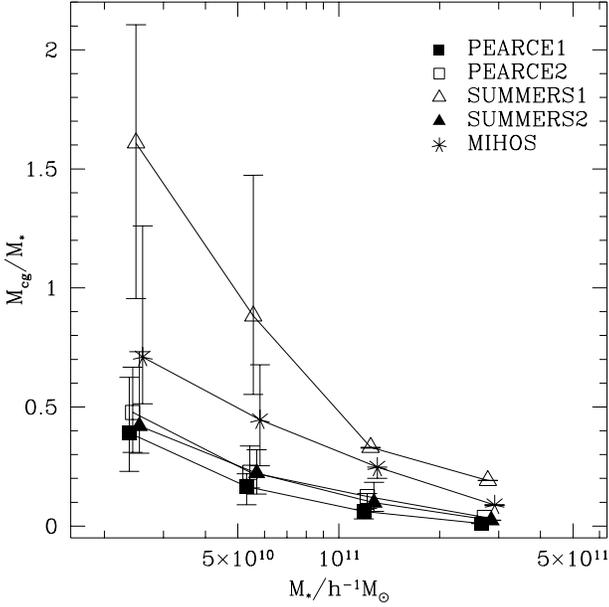,height=8.5cm}}
\caption{The median ratio of cold gas to stars at $z=0$, plotted against
the stellar mass of each galaxy with non-zero gas content, for simulations 
with star formation. Error bars represent the 10 and 90 percentile points of 
the distribution in each bin.}
\label{fig:cgf}
\end{figure}

In Fig.~\ref{fig:cgf}, we plot the median mass ratio of cold gas to
stars against the stellar mass of each galaxy with 32 or more star
particles and a non-zero gas mass, in bins of width, $\Delta \log M_*=0.35$. 
Error bars represent 10 and 90 percentile points of the distribution in each bin. 
There is a clear trend for more massive galaxies to contain proportionally 
less cold gas. This trend reflects the distribution of mean stellar ages
with galaxy mass: the more massive galaxies in all star formation models
harbour, on average,  older stellar populations than less massive galaxies.
Therefore, cooled baryons have had more time to collapse and cross the
imposed threshold for star formation.

The cold gas to star mass fraction in the brightest galaxies
(above $M_*=10^{11} \hMsol$, approximately the mass of an $L^*$ 
galaxy) is less sensitive to the choice of star formation prescription
than for less massive galaxies. However, below $M_*=10^{11} \hMsol$, 
simulations with a higher density threshold (e.g. {\sc summers2}) contain 
significantly more cold gas than simulations with a lower threshold. For 
example, the median ratio of cold gas to stars at $M_*=2.4\times 10^{10}\hMsol$ 
is $\sim 0.4$ for {\sc summers2} but is 1.6 for {\sc summers1}. Again, this 
trend reflects the reduction in star formation activity when the density
threshold is high.

\begin{figure}
\centering
\centerline{\psfig{file=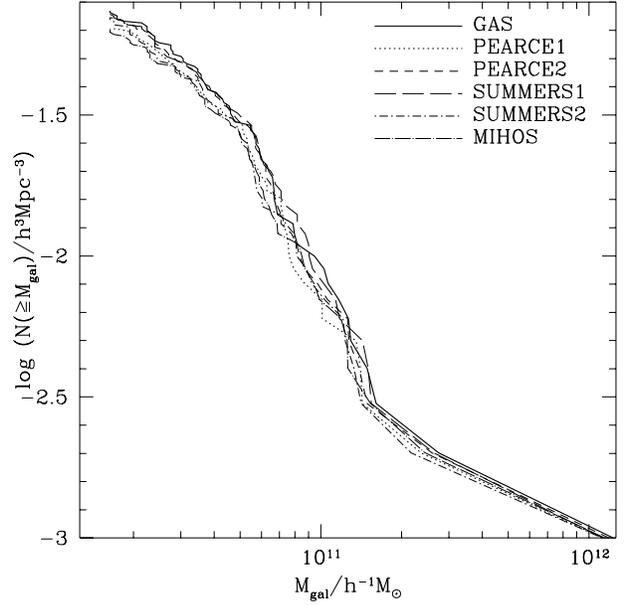,height=8.5cm}}
\caption{Cumulative mass functions for galaxies 
in simulations with star formation at $z=0$. The galaxy mass includes
cold gas and stars. For comparison, the mass function for {\sc gas} is
also plotted.}
\label{fig:mf}
\end{figure}

Observations of atomic and molecular gas in disk galaxies can,
in principle, be used to place constraints on plausible star formation 
models. However, this can only be done in a rather crude manner because 
the simulations do not have enough resolution to distinguish between 
morphological types. By selecting those galaxies with cold gas
we are at least able to identify candidate disk galaxies in the 
simulations. From the data plotted in Fig.~9 of Cole \etal 
(\shortcite{Cole2000}), $L^*$ disk galaxies ($M/L_B \sim 2$
and $M_B^* \sim -19.5$) have $M_{\rm cg}/M_* \sim 0.1-0.3$, values
which do not increase significantly for galaxies an order of magnitude
less massive. This is broadly consistent with our results for
simulations with low density thresholds, but inconsistent with those
with high thresholds (extrapolating the result for {\sc summers1} down
to $M_*=10^{10}\hMsol$ implies a ratio of at least 2).

\subsubsection{Galaxy mass and luminosity functions}
\label{subsubsec:gal_lf}

Cumulative mass functions for the galaxy population at $z=0$ are
plotted in Fig.~\ref{fig:mf}.  For comparison, the mass function for 
simulation {\sc gas} is also displayed. All the mass
functions are very similar, demonstrating that the abundance of
galaxies over the mass range resolved in these simulations is
insensitive to the inclusion of star formation.

\begin{figure}
\centering
\centerline{\psfig{file=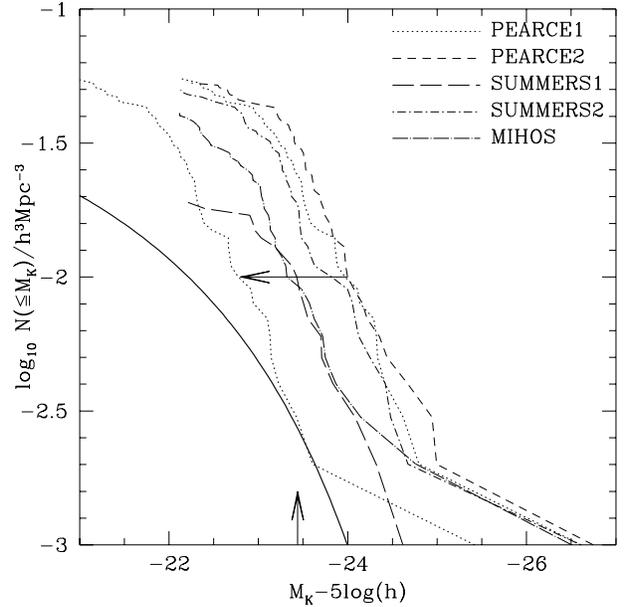,height=8.5cm}}
\caption{Cumulative $K$--band luminosity functions of the galaxies at
$z=0$ for simulations with star formation. The Schechter fit to the
luminosity function from the 2dF and 2-Mass surveys (Cole \etal 2001)
is also plotted (solid curve), with the value of $M_{K}^*$ shown as a
vertical arrow. The horizontal arrow indicates the shift that would be
required for simulation {\sc pearce1} to agree with the observational
data at $M_{K}^*$}
\label{fig:lf}
\end{figure}

We can assign luminosities to the galaxies in the simulations using a
stellar population synthesis model. We calculate $K$--band
luminosities since the light in this band is less affected by dust
than at other bands and, since it is dominated by late--type stars, it
provides a more robust estimate of the underlying stellar mass of the
galaxy. To calculate galaxy luminosities, the formation time, $t_f$,
of each star particle, $m_*$, was used to calculate the luminosity per
unit mass at the present day, $l(t_0-t_f)$. The total magnitude is
then
\begin{equation}
M_{K} = 5\log h + M_{K}(0)-2.5 \, \log \sum_{i=1}^{N_{*}} \, 
l(t_0-t_f) \, m_*,
\label{eqn:mag}
\end{equation}
where $M_{K}(0)$ is the zero--point and $N_{*}$ is the number of
star particles in the galaxy. The function, $l(t_0-t_f)$ was
generated using the stellar population synthesis models of Bruzual \&
Charlot (\shortcite{BC93}), assuming solar metallicity and the stellar
IMF proposed by Kennicutt (\shortcite{Kennicutt83}; see also
\cite{Cole2000}). The total magnitude of each galaxy can be
rescaled to take into account non--visible stellar material 
(\ie a population of brown dwarfs) by adding the term 
$2.5\log \Upsilon$, where $\Upsilon \equiv 1+M({\rm brown \,
dwarfs})/M({\rm visible \, stars})$ (\cite{Cole2000}). 

Cumulative luminosity functions for galaxies in our simulations at $z=0$
are shown in Fig.~\ref{fig:lf}.  Only galaxies with 32 or more star
particles (or equivalent) are included. The best--fit Schechter function to
the $K$--band galaxy luminosity function from the 2dF and 2-Mass surveys is also
plotted as a solid curve (\cite{Cole2001}), with the value of $M_{K}^*$
marked with a vertical arrow. The model luminosity functions are plotted
assuming $\Upsilon=1$; the shift required for simulation {\sc pearce1} to agree
with the data at $M_{K}^*$ is indicated by a horizontal arrow. This
implies a value of $\Upsilon=3$, \ie twice as much mass is required in brown dwarfs 
than in visible stars, which is in clear contradiction with observations
(see \cite{Balogh2001}).

\subsection{Summary}
\label{subsec:sf_summary}

We have examined various prescriptions in the literature for turning
gas into stars in cosmological simulations. The determinant factor in
each model is the range of density (usually defined by a lower limit,
or threshold) at which star formation is allowed to proceed
efficiently: simulations with lower thresholds naturally form more
stars than simulations with higher thresholds. The cooling rate of the
gas is largely unaffected by the inclusion of a star formation
mechanism.  In these models, the star formation efficiency determines
the fraction of galactic baryons that end up in the form of cold gas
or stars. Simulations with a high density threshold contain more cold
gas in (particularly sub-$L_*$) galaxies than simulations with low density
thresholds; the former are discrepant with observational data.  All of
the models overpredict the abundance of galaxies at a given magnitude,
or equivalently produce galaxies that are (about 0.5-1 magnitudes) too
bright at a given abundance. In the next section, we consider feedback
mechanisms that act to reduce galaxy luminosities.

\section{Feedback}
\label{sec:sf_feedback}

In this section we study the effects of energy injection associated 
with Type~II supernovae explosions in a cosmological
simulation. 
On energetic grounds alone, these events must be important
during galaxy formation, and their role as a negative feedback
mechanism allows {\it ab-initio} hierarchical models to match the
observed galaxy luminosity function (\cite{WR}; \cite{WF};
\cite{Cole91}). Ideally, we would like to be able to resolve
individual supernova events and model the multi-phase intergalactic
medium in our simulations. However, this is impractical, both because
of computational limitations and because of the lack of a detailed
understanding of the processes involved. At present, the only
practical strategy is to include the effects of supernova feedback
phenomenologically.

Phenomenological feedback models within numerical simulations have
already been considered by a number of authors studying galaxy
formation/evolution.  For example, supernova feedback was first
included within SPH simulations by Katz(\shortcite{K92}), who studied
the collapse of a galaxy-sized density perturbation, and used a scheme
where every star particle supplied neighbouring gas particles with
thermal energy (heat). However, this had very little effect on the
galaxy's properties, because the high gas density at the locations
where energy is injected implies a cooling time which is much shorter
than the dynamical time. As a result, the energy injected is simply
radiated away. For future reference we will refer to this scheme for
energy injection as ``thermal feedback''.

An alternative feedback scheme was proposed by Navarro \& White
(\shortcite{NW93}). In this model, a fraction, $f_v$, of supernovae
energy is supplied to neighbouring gas particles in the form of
kinetic energy, with the rest being added as heat.  Like Katz, Navarro
\& White found that for low values of $f_v$ (\ie almost all thermal energy),
very little feedback occurs and consequently the star formation rate
is very high. However, for higher values of $f_v$ the star formation
rate is reduced because kinetic perturbations are able partially to
halt the local collapse of a density perturbation.  We refer to this
method as ``kinetic feedback''. Kinetic feedback has since been used
by other authors (\eg \cite{Mihos&Hernquist}).

Subsequent attempts have been made to boost the effects of thermal
feedback by preventing the gas from immediately re-radiating the
supernova energy. This was first considered by Gerritsen
(\shortcite{G97}) who argued that previous feedback schemes did not
produce an overall disk morphology resembling those observed,
with spherical cavities, or bubbles, filled with hot gas. They found
that injecting all the energy into one particle as heat and
preventing it from cooling for a finite length of time, had the
desired effect. A more elegant approach was taken by Springel
(\shortcite{Springel2000}), who created a second pressure term in the
equations of motion, representing turbulent pressure caused by
supernovae. Yet another example is the study by Thacker
\& Couchman (\shortcite{Thacker&Couchman2000}), who considered a scheme
in which a ``cooling'' density for each gas particle injected with
energy is obtained from a simple pressure equilibrium argument. This
density is lower than the standard SPH density, and so lengthens the
particle's cooling time, before resorting back to the SPH density
after a finite period.

Attempts to simulate galaxy formation in a large region, starting from
realistic cosmological initial conditions, and including a model for
feedback have been made by several groups using Eulerian grid codes
(\eg \cite{CO92}, \shortcite{CO99}; \cite{YKKK}) and SPH codes
(\cite{Summers_phd}; \cite{KWH96}). The former technique is hindered
by poor spatial resolution, with galaxies being several factors
smaller than the minimum resolved scales (although this situation is
improving, particularly with adaptive mesh codes). Studies using SPH
simulations have focussed on thermal feedback, and like Katz
(\shortcite{K92}), have also found that adding thermal energy to the
gas makes little difference to the resulting galaxy population.  In
the next subsection we will describe our own implementations of
thermal and kinetic feedback in our simulations, which we regard as a
natural follow-up to the work of Summers (\shortcite{Summers_phd}) and
Katz \etal (\shortcite{KWH96}).  As we shall see, our models are
capable of producing strong feedback, reducing the amount of material
which cools and forms stars, and resulting in a more realistic
$K$-band luminosity function.

\subsection{Simulation details}

\subsubsection{Cooling and Star formation}

We have studied the effects of feedback in the $\Lambda$CDM and SCDM 
cosmological models described in \S~\ref{subsec:cosmology}.  Similar
trends with respect to variation of the parameters are present in both
models, so we will focus mainly on $\Lambda$CDM. We implement the
``decoupling'' technique described by Pearce \etal
(\shortcite{Pearce2000}) whereby gas particles with temperature below
$T=12,000\kelvin$ are ignored when estimating the gas density of hot
particles with temperature above $T=10^{5}\kelvin$. This technique
effectively prevents the unphysical cooling of hot gas induced purely
by the presence of nearby cold, dense galactic gas. (Note, however, 
that this problem is largely circumvented due to the inclusion of a 
star formation prescription which removes the majority of galactic 
material from the SPH calculations.) This
modification is a crude attempt to model a multiphase gas with SPH and
has been discussed further elsewhere (\eg K2000; \cite{Croft2000}; see
also \cite{Ritchie&Thomas}). 

Star formation was incorporated using the Pearce method (\cf
\S~\ref{subsubsec:pearce}), assuming an overdensity threshold of 5000
and a temperature threshold of $30,000\kelvin$ (essentially the {\sc
pearce1} criteria). When a gas particle satisfies both criteria it is
instantaneously converted into a star particle.  In the $\Lambda$CDM
simulations we assumed an unevolving gas metallicity of $Z=0.3\Zsol$
while for the SCDM simulations, we retained the metallicity of
$Z=0.5\Zsol$ assumed in the previous section. Note that our choice of
metallicity results in an overestimate of the cooling rate of gas at
high redshift, where $Z$ is considerably smaller. Self-consistent
models of feedback and metal enrichment within simulations will be
presented in a future paper.

\subsubsection{Supernova energetics}

We adopt the same parameter values for the supernovae energetics as 
Katz (1992) and Katz \etal (\shortcite{KWH96}). To summarise, the 
amount of energy available from supernovae is calculated assuming 
that stars with mass above $8 \Msol$ release $10^{51}\erg$ of energy 
into the surrounding gas. A Miller \& Scalo (\shortcite{Miller&Scalo79}) IMF
is assumed, with lower and upper mass limits of
$0.1\Msol$ and $100\Msol$ respectively, implying a specific
energy released in supernovae of
$\sim 3.7 \times 10^{15} \erg \,{\rm g}^{-1}$, equivalent to a temperature of 
$1.8 \times 10^{7} \kelvin$.
%\footnote {{\bf *** Scott: do you mean E=kt here?} $T=(\gamma-1)\mu m_{\rm p}/k$, where $\gamma=5/3$ is the ratio of 
%specific heats at constant volume and constant pressure, for a
%monatomic ideal gas; we assume $\mu=0.59$, the mean molecular weight
%of a gas of primordial composition}.
The size of the timestep in the simulations (typically $3 \,{\rm Myr}$)
is about one order of magnitude smaller than the lifetime of an
$8\Msol$ star ($\tau_8 \sim 20 {\rm Myr}$). For this reason, the
energy was released gradually, using the following function
\begin{equation}
\dot{u}_{\rm SN}(t) = 3.7 \times 10^{15} \, \exp \, ( - (t-t_0) / \tau_8 ) 
\, \tau_8^{-1} \, \erg \, {\rm g^{-1}} ,
\label{eqn:snenergy}
\end{equation}
where $t_0$ is the time when the star formation event occurred. Energy was
released until $t-t_0=200 \Myr$, when $\dot{u}_{\rm SN}$
becomes negligible. 

\subsubsection{Feedback models}

Our prescription for thermal feedback works as follows.  When a gas
particle is to be converted into a star particle, it first passes through
an intermediate phase, in which it is labelled a {\it supernova}
particle. Supernova particles feel the gravitational force only, but the
SPH algorithm maintains both their densities and neighbour list. Smoothing
lengths are fixed at the minimum value, $h_{\rm min}$, and each gas
neighbour within the SPH smoothing radius (2$h_{\rm min}$) receives a
proportion of the thermal energy available at that time. For supernova
particle $i$, the energy given to gas particle $j$, over the timestep
$\Delta t$, is
\begin{equation}
\Delta u_{{\rm SN},j} = \epsilon \, \dot{u}_{{\rm SN},i} \Delta t \, 
                        {H(2h_{\rm min}-|{\bf r}_i-{\bf r}_j|) \,
                        \over
                        \sum_{j=1}^{N_{\rm gas}} \, 
                        H(2h_{\rm min}-|{\bf r}_i-{\bf r}_j|)} \,,
\label{eqn:fb_thermal}
\end{equation}
where $0 \leq \epsilon \leq 1$ is the feedback efficiency parameter, and
$H$ is the Heaviside step function. We choose not to weight the
contributions using the SPH kernel since this as an unnecessary
complication in cosmological simulations, where the forces within galaxies
are softened. Furthermore, the efficiency of energy transfer will vary
depending on the number of neighbours of each supernova particle. Our
method ensures that the same amount of energy per supernova is always
transferred to the gas and so the efficiency is controlled solely by the
choice of $\epsilon$.

As discussed above, this scheme is expected to have little effect on
the galaxy properties, since the gas will re-radiate the energy in a
short time.  To circumvent this problem, we follow the method of
Gerritsen (\shortcite{G97}) and prevent the gas from cooling over a
period of time. This timescale is controlled using the parameter
$\tau$, which is the length of the non-radiative period in units of
the age of the universe at $z=0$. 

We also study kinetic feedback, where instead of heating the gas, it
receives a velocity perturbation radially away from the supernova 
particles. The total magnitude of the velocity added to gas particle $j$,
from supernova particle $i$, is
\begin{equation}
\Delta v_{\rm SN,j} = \sqrt{2 \, \epsilon \, \dot{u}_{{\rm SN},i} \, \Delta t
  \, {H(2h_{\rm min}-|{\bf r}_i-{\bf r}_j|) \,
    \over
    \sum_{j=1}^{N_{\rm gas}} \, H(2h_{\rm min}-|{\bf r}_i-{\bf r}_j|)}}.
\label{eqn:snkinenergy}
\end{equation}
This scheme is the ``momentum'' implementation of \cite{NW93}.
%For kinetic feedback, the problem of instantaneous re--radiation is not
%apparent, due to the failure of the simulations to resolve shocks (which
%would efficiently thermalize the gas) on galactic scales (\cite{KWH96}). 
%Hence, we set $\tau=0$, \ie the cooling rate of the gas is limited by the 
%timestep.

\subsection{Results}
\label{subsec:sf_feed_results}

\begin{table}
\begin{center}
\caption{General properties at $z=0$ of the simulations with feedback}
\begin{tabular}{cllllr}  
Cosmology  & Feedback & $\epsilon$ & $\tau$ & $f_{\rm gal}$ & $N_{\rm gal}$\\
\hline
$\Lambda$CDM & None     & N/A   & N/A  &  0.35    & 34\\
SCDM         & None     & N/A   & N/A  &  0.26    & 53\\
\\                                              
$\Lambda$CDM & Thermal  & 0.1   & 0.01 &  0.28    & 30\\
$\Lambda$CDM & Thermal  & 0.1   & 0.1  &  0.17    & 26\\
$\Lambda$CDM & Thermal  & 0.1   & 1.0  &  0.096   & 17\\
$\Lambda$CDM & Thermal  & 1.0   & 0.01 &  0.11    & 13\\
$\Lambda$CDM & Thermal  & 1.0   & 0.1  &  0.083   & 12\\
$\Lambda$CDM & Thermal  & 1.0   & 1.0  &  0.065   & 13\\
\\                                              
SCDM         & Thermal  & 0.1   & 0.01 &  0.19    & 49\\
SCDM         & Thermal  & 0.1   & 0.1  &  0.099   & 35\\
SCDM         & Thermal  & 0.1   & 1.0  &  0.061   & 21\\
SCDM         & Thermal  & 1.0   & 0.01 &  0.076   & 24\\
SCDM         & Thermal  & 1.0   & 0.1  &  0.047   & 14\\
SCDM         & Thermal  & 1.0   & 1.0  &  0.037   & 11\\
\\                                              
$\Lambda$CDM & Kinetic  & 0.001 & N/A  &  0.34    & 29\\
$\Lambda$CDM & Kinetic  & 0.01  & N/A  &  0.22    & 20\\
$\Lambda$CDM & Kinetic  & 0.1   & N/A  &  0.056   & 11\\
\\                                              
SCDM         & Kinetic  & 0.001 & N/A  &  0.23    & 52\\
SCDM         & Kinetic  & 0.01  & N/A  &  0.15    & 35\\
SCDM         & Kinetic  & 0.1   & N/A  &  0.046   & 11\\
\end{tabular}
\label{tab:feed_details}
\end{center}
\end{table}

We present results from 6 different parameter choices for the thermal
feedback model and 3 for the kinetic model.  Specifically, for
simulations with thermal feedback, we consider $\tau={0.01,0.1,1.0}$
for $\epsilon={0.1,1.0}$. For simulations with kinetic feedback, we
consider $\epsilon={0.001,0.01,0.1}$. Details of the feedback models
for $\Lambda$CDM and SCDM simulations are summarised in
Table~\ref{tab:feed_details}.

%Table~\ref{tab:feed_details} gives
%properties of SCDM and $\Lambda$CDM simulations with feedback and, 
%for comparison, also for the corresponding simulations without feedback. 
%The first two columns list the cosmological model and feedback model respectively,
%while Columns 3 \& 4 give the values of $\epsilon$ and $\tau$.

\subsubsection{General properties at $z=0$}
\label{subsubsec:sf_feed_gen}

We identify galaxies by applying a friends-of-friends group-finder to the
gas and star particles as in Section~\ref{subsubsec:sf_stars_gen}, except
that we use a linking length $b=0.2$. This gives a slightly larger number
of galaxies than $b=0.1$. (In simulations without feedback, both values of
$b$ pick out the same galaxies, but when feedback is included a larger
linking length is required because the galaxy material becomes slightly
distended by the disturbance produced by the injection of supernova energy.
Since the internal structure of galaxies is unresolved, the exact choice of
linking length is unimportant provided it captures the majority of galaxy
material.) We also demand that each galaxy should contain a minimum of 32
star particles, regardless of the number of gas particles, in order to
provide a reasonable estimate of its luminosity.  The minimum baryonic mass
of a galaxy in these simulations is $1.0
\times 10^{10} \hMsol$ for the $\Lambda$CDM model and $1.6\times 10^{10}
\hMsol$ for the SCDM model. 

Column~5 of Table~\ref{tab:feed_details} gives the fraction of baryons
(cold gas and stars) in galaxies, labelled $f_{\rm gal}$.  For simulations
with the same feedback parameters, the fraction of galaxy material is
always higher in the $\Lambda$CDM case than in the SCDM case though the
trends with varying $\epsilon$ and $\tau$ are the same. A larger feedback
efficiency, $\epsilon$, causes a decrease in the fraction of galactic
material at $z=0$ (for both thermal and kinetic feedback). In the thermal
case, increasing the efficiency increases the pressure of reheated gas
particles, with the result that more material can rise out of galaxies,
before it can recool and form stars. In the kinetic case, a larger
$\epsilon$ causes an increase in the fraction of gas particles with
velocities greater than the escape velocity of each galaxy.  For a given
value of $\epsilon$ in the simulations with thermal feedback, a larger
$\tau$ results in a smaller $f_{\rm gal}$. This is because the longer the
reheated gas is prevented from cooling, the easier it is for this gas to
escape from the galactic potential. (Note that for the largest value of
$\tau$ considered, $\tau=1$, reheated gas can never re-radiate its energy
before the end of the simulation.)

The number of resolved galaxies in each simulation at $z=0$ is listed
in Column~6. This number is sensitive to the feedback mechanism.  In
general, SCDM simulations contain more galaxies than the equivalent
$\Lambda$CDM simulations, although the exact values depend on both the
mass resolution of the simulation and the shape of the galaxy
mass function near this threshold.
 
\subsubsection{Star formation rates}
\label{subsubsec:sf_feed_sfr}

\begin{figure}
\centering
\centerline{\psfig{file=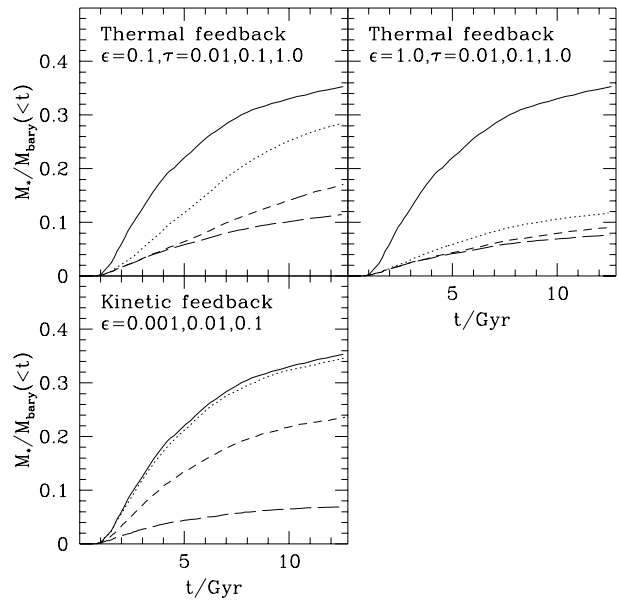,height=8.5cm}}
\caption{Cumulative fraction of baryons in stars plotted 
against time, in Gyr, for $\Lambda$CDM simulations without feedback
(solid lines) and with feedback. For thermal feedback, dotted lines
correspond to $\tau=0.01$, short-dashed to 
$\tau=0.1$ and long-dashed to $\tau=1$; for kinetic feedback, the
same sequence corresponds to $\epsilon=0.001,0.01$ \& 0.1. 
Model parameters are given in the legend in each panel.}
\label{fig:fsfr}
\end{figure}

In Fig.~\ref{fig:fsfr}, we illustrate the effect of feedback on the
global star formation rate, by plotting the cumulative fraction of
baryons in stars as a function of time for $\Lambda$CDM models. In
each panel we compare results between simulations with the same
feedback prescription, but with different values of one parameter.
For thermal feedback, dotted lines represent results from simulations
with $\tau=0.01$, short-dashed with $\tau=0.1$ and long-dashed with
$\tau=1$; for kinetic feedback, the same sequence represents results
for $\epsilon=0.001,0.01$ \& 0.1. The equivalent simulation without
feedback is also plotted as a solid line.

Consider first the case of thermal feedback. For $\epsilon=0.1$, the
efficiency of star formation is sensitive to the value of $\tau$ (when
varied over 2 orders of magnitude). However, increasing $\epsilon$ by
a factor of 10 produces only small differences in the star formation
rate when $\tau$ is varied. In the latter case, enough energy is
deposited into the gas that even on the shortest timescale considered,
$\tau=0.01$ (of order 100 Myr), much of the reheated material is
unable to cool and form stars before the end of the simulation.
Kinetic feedback can be very effective at reducing the amount of star
formation, even though gas above $10^{4}\kelvin$ is able to cool
radiatively at all times. Although supplying 0.1 per cent of the
available energy makes no significant difference to the star formation
rate, increasing the efficiency to 1 and 10 per cent lowers the
fraction of stars formed by $z=0$ to around 2/3 and 1/4 of the
no-feedback values respectively.

\subsubsection{The fate of reheated gas}

\begin{figure}
\centering
\centerline{\psfig{file=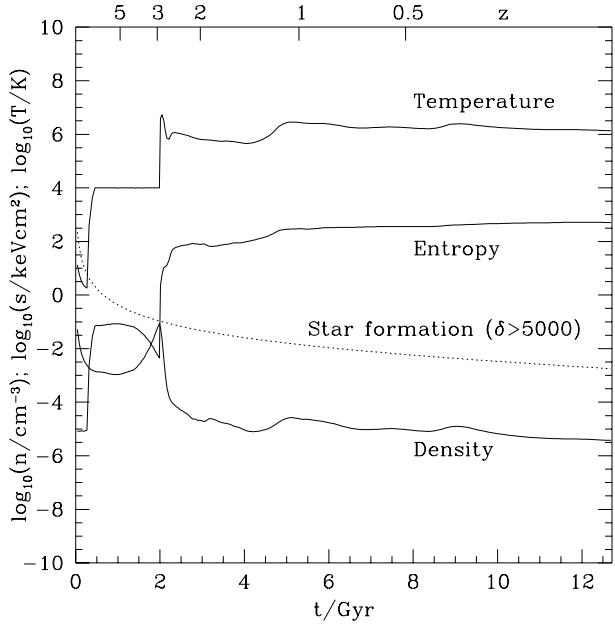,height=8.5cm}}
\caption{Thermal history (density, temperature and entropy as a
function of time) of a gas particle in the $\Lambda$CDM simulation
with thermal feedback ($\epsilon=1, \tau=1$) that would otherwise have
cooled and formed stars. The density threshold for star formation is
plotted as a dotted line. Redshift intervals are labelled along the
top of the figure.}
\label{fig:track}
\end{figure}

Virtually all the gas that is reheated, either thermally or kinetically,
has a temperature in the range $10^{4}-10^{5}\kelvin$. The temperature
to which it is reheated and its subsequent thermal evolution depends
on the specific feedback mechanism. For thermal feedback, at least
80\% of the remaining gas that was once reheated is hotter than
$10^5\kelvin$ since any gas that can cool again is rapidly turned into
stars. In the case of kinetic feedback, the final temperature of the
reheated gas depends on the value of $\epsilon$: a greater efficiency 
leads to hotter gas. Unlike in the thermal models, gas in the kinetic
models is not reheated instantaneously but has to thermalise its
energy, a process which will be particularly susceptible to numerical 
resolution (Katz \etal \shortcite{KWH96}).

To investigate the fate of reheated gas, we first illustrate the
typical history of a reheated gas particle that would otherwise have
cooled within a galaxy and been converted into a star particle. We
have selected a particle in the $\Lambda$CDM simulation with thermal
feedback, for $\epsilon=1$ and $\tau=1$. The density ($n/\icm3$),
specific ``entropy'' (defined as $s=kT/n^{2/3}$) and temperature of this
particle are plotted in Fig.~\ref{fig:track}. Also shown (dotted line)
is the density at each redshift equivalent to the overdensity
threshold for star formation, \ie $n > 5000 \, \left<\rho\right>/\mu m_{\rm H}$. 

Initially ($z>5$), the particle cools adiabatically (\ie isentropically) 
from its initial temperature of 100K due to the expansion of the Universe. 
(Note that our choice of initial temperature is arbitrary but is too low
to have any effect on the subsequent evolution of the particle. The true
temperature of the gas at high redshift is determined by the 
intensity of the UV background, which we ignore for this study.)
At $z=5$ the particle has fallen into a halo where it
is shock-heated to a temperature of $10^{4}\kelvin$: the density is
sufficiently high that the particle is able to radiate any excess 
thermal energy within a timestep (K2000). At this point the density
does not increase (as expected during shock-heating) since the region 
is not fully resolved by the SPH algorithm. A few of the particle's
neighbours are not part of the overall collapse (leading to an 
underestimate of the gas density) even though the net flow is converging, 
which results in viscous heating. Shortly after this time, however, the 
density of the particle starts increasing as it condenses to form part of 
a ``galaxy'' within a forming dark matter halo. (The halo, with 
$V_c \sim 60 \kms$, contains approximately 25 particles and so is also
poorly resolved at this stage.) The density reaches a peak value of 
$0.1 \icm3$ at $z \sim 3$. However, before the particle crosses the threshold for 
star formation, it absorbs supernovae energy which raises its temperature by 
about 2 orders of magnitude, to $10^{6}\kelvin$.  Correspondingly, the density 
decreases by 3 orders of magnitude (aided by the fact that since the
particle is at $T>10^{5}\kelvin$, its SPH density is determined only
by other hot particles) and its entropy jumps to $\sim 100 {\rm keV
cm}^{2}$. Apart from a few minor heating events, the particle evolves
approximately isentropically for the remaining time ($z<3$) as it moves
into less dense environments.

\begin{figure*}
\centering
\centerline{\psfig{file=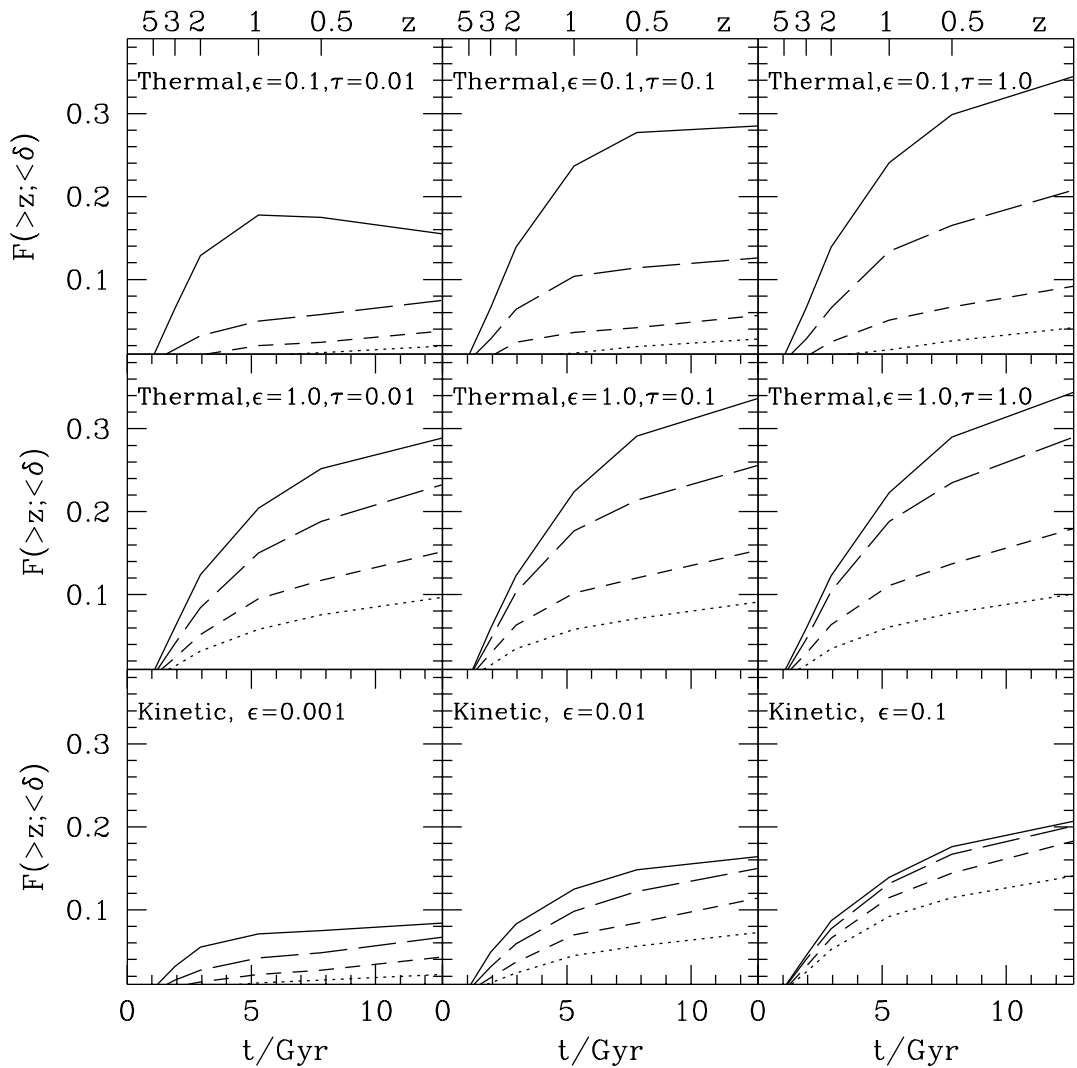,height=15cm}}
\caption{The fraction of gas reheated by supernovae energy as a
function of time, in Gyr, lying in regions of different overdensity. Solid
lines illustrate the total reheated fraction, long-dashed lines the
fraction with $\delta<500$, short-dashed lines the fraction
with $\delta<50$ and dotted lines the fraction with $\delta<10$.
Redshift intervals are labelled along the top of the figure.}
\label{fig:deltaz}
\end{figure*}

In order to quantify the fate of gas reheated by supernovae energy, we
plot in Fig.~\ref{fig:deltaz} the fraction of reheated gas particles
in the $\Lambda$CDM simulations as a function of time, which lie
in regions of different overdensity. The solid lines show the entire
reheated fraction, the long-dashed lines the fraction with
$\delta<500$, the short-dashed lines the fraction with $\delta<50$ and
the dotted lines the fraction with $\delta<10$. The first cut
approximately selects gas that has been expelled from galaxies ({\it
blow-out}), the intermediate overdensity preferentially picks out gas
expelled from haloes altogether ({\it blow-away}), while the final
criterion indicates particles that have been ejected into weakly
overdense regions such as those responsible for the Ly-$\alpha$ forest
observed in quasar spectra.

The first reheated particles appear at around $z=5$, when the first
stars form and the associated supernovae inject energy into the
gas. The reheated fraction increases steadily to $z=0$ in all
simulations except in the case of thermal feedback with 
$\epsilon=0.1$ and $\tau=0.01$.  In this simulation, the amount of
reheated gas that can cool again and form stars at $z<1$ outweighs the
total amount of gas that is reheated. As expected, at low redshift,
the feedback models with the highest star formation rates have the
lowest fraction of reheated gas. However, the amount of reheated
material is lower in the kinetic feedback models at any given time,
than in a thermal model which produces approximately the same fraction
of stars. For example, both the thermal feedback model with
$\epsilon=1,\tau=1$ and the kinetic feedback model with $\epsilon=0.1$
predict that 5-6\% of the baryons are locked into stars in resolved
galaxies ($M>10^{10}\Msol$) at $z=0$. However, $\sim 34$\% of the
baryons are reheated in the thermal case, whereas only $\sim
20$\% are reheated in the kinetic case.

In all our simulations, a significant fraction of the gas that absorbs
supernova energy escapes from galaxies and haloes. Kinetic feedback is
more efficient at transporting the gas into the intergalactic medium
than thermal feedback.  In the latter models, as much as 10\% of the
baryons are reheated and expelled into weakly overdense regions
($\delta<10$) by $z=0$, so that the ratio of diffuse reprocessed gas
to stars is as high as 2.  Kinetic feedback models predict a similar
ratio, even though the total fraction of reheated gas is much lower.

\subsubsection{Galactic winds}

\begin{figure}
\centering
\centerline{\psfig{file=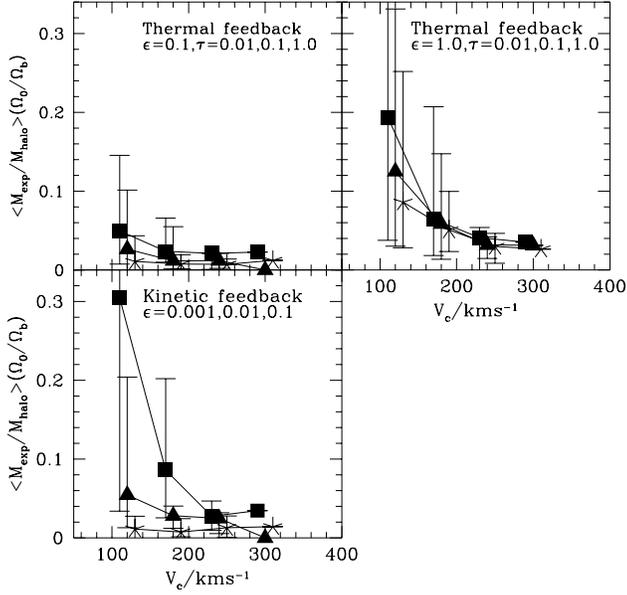,height=8.5cm}}
\caption{The median mass fraction of gas expelled from haloes as
a function of the circular velocity of the halo from which it was
expelled. Error bars illustrate 10 and 90 percentile points 
within each bin.
Stars, triangles and squares represent thermal feedback
simulations with $\tau=0.01,0.1$ \& 1.0, and kinetic feedback
simulations with $\epsilon=0.001,0.01$ \& 0.1 respectively.}
\label{fig:expel}
\end{figure}

One of the effects of energy injection is to expel gas from haloes in
the form of a galactic wind.  The time at which the gas is expelled is
only approximately known in the simulations because of the relatively
sparse output times. To estimate it, we first identify dark matter
haloes at the simulation output times ($z=5,3,2,1,0.5,0$). Haloes are
defined as overdense spheres enclosing an average density $\Delta_c \,
\rho_{\rm cr}(z)$, where $\rho_{\rm cr}(z)=3H^{2}(z)/8\pi G$; we set
$\Delta_c=178 \, \Omega^{0.45}(z)$ for the $\Lambda$CDM model
(\cite{ECF}). We then tag each particle with the latest output time when it
was inside a halo, and note the corresponding halo mass. The circular velocity
of a halo is related to its mass and redshift by
\begin{equation}
V_c =  740 \, M_{14}^{1/3} \, 
              \left({H(z) \over H_0}\right)^{1/3} \, 
              \left({\Delta_c \over 178}\right)^{1/6} \, \kms,
\label{eqn:vcirc}
\end{equation}
where $M_{14}$ is the total mass of the halo in $10^{14}\hMsol$. A
reheated gas particle is deemed to have been expelled from the halo
when it first moves beyond the virial radius, providing it was
identified with a resolved halo (32 or more particles) before the reheating 
event and also prior to expulsion. All expelled particles are then grouped 
together to calculate the expulsion mass for each halo. 

In Fig.~\ref{fig:expel}, we plot the median mass fraction of expelled
gas as a function of the circular velocity of the halo from which the
gas was blown away. (We define the halo gas mass as $M_{\rm halo}
\, \Omegab/\Omega_0$, where $M_{\rm halo}$ is the total halo mass.)
Error bars illustrate 10 and 90 percentile points of the distribution
within each bin.
Stars, triangles and squares represent results from $\Lambda$CDM
thermal feedback simulations with $\tau=0.01,0.1$ \& 1.0, and kinetic
feedback simulations with $\epsilon=0.001,0.01$ \& 0.1 respectively.
Although the range of resolved circular velocities is limited in these
simulations, Fig.~\ref{fig:expel} shows a clear trend of decreasing
expulsion fraction with increasing circular velocity for nearly all 
cases, with only a few per cent of gas escaping from Milky Way sized haloes 
($V_c \sim 220 \kms$) and larger.
Increasing the available energy (large $\epsilon$) causes a larger
fraction of the gas to be expelled, particularly in the haloes with the 
lowest values of $V_c$. A larger value of $\epsilon$ raises the mean 
temperature of the reheated gas, allowing a greater fraction to 
escape from the smaller systems. However, the average cooling times for 
these systems are considerably shorter than their dynamical times, and so 
the amount of gas escaping in the thermal feedback simulations is
lower for smaller values of the delay timescale, $\tau$.

\subsubsection{Galaxy properties}
\label{subsubsec:sf_feed_galmass}

\begin{figure}
\centering
\centerline{\psfig{file=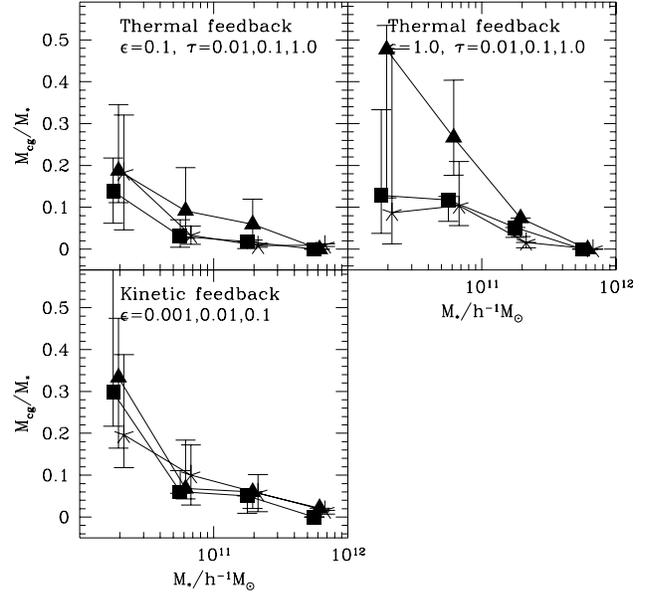,height=8.5cm}}
\caption{The $z=0$ median mass ratio of cold gas to stars plotted
against the stellar  
mass of each galaxy. Error bars represent 10 and 90 percentile points within each 
bin. Stars, triangles and squares illustrate results for thermal feedback 
simulations with $\tau=0.01,0.1$ \& 1.0, and kinetic feedback simulations with 
$\epsilon=0.001,0.01$ \& 0.1 respectively.}
\label{fig:fcgf}
\end{figure}

We now investigate the effects of feedback on the present day galaxy
population. Firstly, we examine its effect on the amount of cold gas
in galaxies, by plotting the mean ratio of cold gas mass to stellar
mass, ${\rm M_{cg}/M_*}$, against $M_*$ for galaxies (with non-zero
gas content) in the
$\Lambda$CDM simulations with 32 or more particles.  The points in
Fig.~\ref{fig:fcgf} represent the median values within bins of width,
$\Delta \log M_{*}=0.5$, and error bars represent 10 and 90 percentile points
within each bin. The trend of decreasing cold gas fraction with galaxy
mass is similar to that seen in Fig.~2 for simulations without
feedback, but all our feedback models predict median cold gas fractions
below 0.5 for galaxies with $M_*>10^{10} \hMsol$. This agrees well
with the observational data summarised by Cole \etal
(\shortcite{Cole2000}; see Section~\ref{subsubsec:gal_cgf})

\begin{figure}
\centering
\centerline{\psfig{file=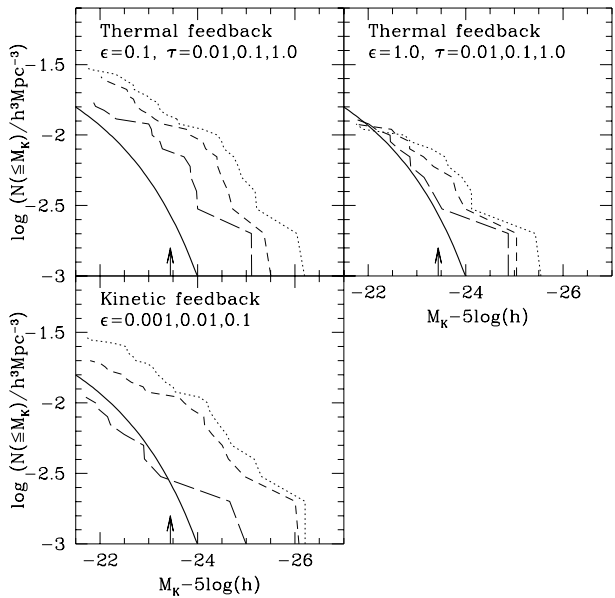,height=8.5cm}}
\caption{$K$--band luminosity function of galaxies in $\Lambda$CDM 
simulations with feedback. Dotted, short-dashed and long-dashed lines 
represent thermal feedback results with $\tau=0.01,0.1$ \& 1.0, and 
kinetic feedback results with $\epsilon=0.001,0.01$ \& 0.1 respectively. 
Also shown (solid curve) is the best--fit Schechter function to the
result from the 2dF and 2-Mass surveys (Cole \etal 2001). The value of 
$M_{K}^*$ is illustrated with a vertical arrow.}
\label{fig:flfk}
\end{figure}

The overall reduction in star formation activity caused by feedback
has a substantial impact on the galaxy luminosity function.
Cumulative $K$--band luminosity functions are shown in
Fig.~\ref{fig:flfk} for the $\Lambda$CDM simulations. Luminosities for
each galaxy (consisting of at least 32 star particles) were calculated
using the method described in Section~\ref{subsubsec:gal_lf}. The best
cumulative Schechter function fit to the data from Cole \etal (2001)
from the 2dF and 2-Mass surveys is plotted as a solid curve, with
the value of $M_K^*$ indicated with a vertical arrow.

Results from the thermal feedback simulation with $\epsilon=0.1,\tau=0.01$
and the kinetic feedback simulation with $\epsilon=0.001$ are similar to
that from a simulation with no feedback. These all produce a luminosity
function which is $\sim 2$ magnitudes brighter than observed.  As the
strength of the feedback is increased (by increasing $\epsilon$ and/or
$\tau$), the stellar mass and hence the luminosity of the galaxies are
reduced.  Both the thermal model with $\epsilon=1$ \& $\tau=1$ and the
kinetic model with $\epsilon=0.1$ predict about the correct number of
$L^*_K$ galaxies, although the latter model underestimates the abundance of
fainter galaxies.

\subsection{Summary}
\label{subsec:f_summary}

We have investigated two ways of including feedback from supernovae energy
in cosmological simulations, in order to reduce the amount of gas that
cools and forms stars. The first method, thermal feedback (\cite{K92}),
adds thermal energy to surrounding gas (with efficiency controlled by the
parameter $\epsilon$) and prevents such material from recooling for a
period (controlled by the parameter $\tau$). The second method, kinetic
feedback (\cite{NW93}), adds velocity perturbations to the gas radially
away from the stars, and allows the gas to cool at all times. Increasing
$\epsilon$ and/or $\tau$ increases the strength of feedback, reducing the
global star formation rate, but without significantly affecting the average
ratio of cold gas to stars in galaxies. The feedback works in both models
by reheating gas which has already cooled, with stronger feedback resulting
in more gas escaping into regions of low overdensity.  The efficiency with
which gas is expelled from dark matter haloes is higher in
haloes with $V_c\sim 100 \kms$ than in haloes with higher circular
velocity. By limiting the star formation rate, feedback lowers the $K$-band
luminosity of galaxies and, with a suitable choice of parameter values, it
is possible to reproduce the bright end of the galaxy luminosity function.

\section{Resolution effects}
\label{sec:sf_resol}

%\begin{table*}
%\begin{center}
%\caption{Details of the simulations with feedback and varying resolution}
%\begin{tabular}{cccrlllllr}
%$N$ & $m_{\rm dark}/\hMsol$ & $m_{\rm bary}/\hMsol$ & $\epsilon_0/\hkpc$ & 
%Feedback & $\epsilon$ & $\tau$ & $f_{*}$ & $f_{\rm gal}$ 
%& $N_{\rm gal}$\\
%\hline
%$2\times 32^3$ & $7.96\times 10^{9}$ & $5.08\times 10^{8}$ & 10.0 &
%Thermal & 1.0 & 1.0 & 0.068 & 0.037 & 11\\
%$2\times 48^3$ & $2.36\times 10^{9}$ & $1.51\times 10^{8}$ & 6.7 &
%Thermal & 1.0 & 1.0 & 0.073 & 0.053 & 41\\
%$2\times 32^3$ & $7.96\times 10^{9}$ & $5.08\times 10^{8}$ & 10.0 &
%Kinetic & 0.1 & N/A & 0.078 & 0.046 & 11\\
%$2\times 48^3$ & $2.36\times 10^{9}$ & $1.51\times 10^{8}$ & 6.7 &
%Kinetic & 0.1 & N/A & 0.078 & 0.054 & 36\\
%\\
%\end{tabular}
%\label{tab:hires}
%\end{center}
%\end{table*} 

% I've simplified the table to focus on results - STK 
\begin{table}
\begin{center}
\caption{Properties of the simulations with feedback and varying resolution}
\begin{tabular}{lllllr}
Resolution & Feedback & $\epsilon$ & $\tau$ & 
$f_{\rm gal}$ & $N_{\rm gal}$\\
\hline
Low-res & Thermal & 1.0 & 1.0  & 0.037 & 11\\
High-res & Thermal & 1.0 & 1.0  & 0.053 & 41\\
Low-res & Kinetic & 0.1 & N/A  & 0.046 & 11\\
High-res & Kinetic & 0.1 & N/A  & 0.054 & 36\\
\\
\end{tabular}
\label{tab:hires}
\end{center}
\end{table} 

The simulations discussed so far have a fixed mass resolution, with
the smallest resolved galaxies having a baryonic mass $\sim 10^{10}
\hMsol$, roughly one tenth that of an $L^*$ galaxy. In the absence
of feedback, resolution effects completely determine the outcome of a
simulation: as smaller objects are resolved, increasingly large
amounts of gas cool, raising both the global fraction of cooled
material as well as the mass of the majority of the galaxies. 
Feedback limits the amount of gas that cools and so it seems plausible 
that an appropriate feedback prescription might lead to results that do not
depend strongly on resolution. In this case, an increase in resolution
would lead to smaller galaxies being resolved without affecting the
properties of the larger objects. 

To address the issue of resolution, we performed two additional
simulations, using both thermal and kinetic feedback models.  For
thermal feedback, we set $\epsilon=1$ and $\tau=0.1$; for kinetic
feedback, we set $\epsilon=0.1$.  The cosmological parameters and
initial conditions were identical to those used for the SCDM model in
\S\ref{sec:sf_starform}, except that the number of gas and dark matter
particles were increased from 32768 ($32^3$) of each to 110592 ($48^3$).
This results in a decrease in the minimum baryonic mass of a resolved
galaxy by about a factor of 3, from $\sim 1.6 \times 10^{10} \hMsol$ to
$\sim 4.8 \times 10^{9} \hMsol$.  When constructing the initial density
field, the phases of the large--scale waves were preserved, so that the
same overall large--scale structure is present in all the simulations. We
also increased the spatial resolution by decreasing the softening length in
proportion to $m^{1/3}$, where $m$ is the dark matter particle mass. This
corresponds to a reduction by a factor of 1.5, for the duration of the
simulation, to a final (Plummer) of $\sim 6.7 \hkpc$. All other parameters
were fixed at the values used in the low resolution simulations.

Table~\ref{tab:hires} compares the properties of the high resolution
simulations with their low resolution counterparts.  The last two columns
give the fraction of baryons in resolved galaxies, $f_{\rm gal}$, and the
total number of galaxies, $N_{\rm gal}$. Galaxies are identified as before,
using a friends-of-friends group-finder on the distributions of gas and
star particles with a linking length of $b=0.2$, and only those with 32 or
more star particles are retained. As expected, an increase in resolution
leads to an increase in both the number of resolved galaxies and the galaxy
baryon fraction.

\begin{figure}
\centering
\centerline{\psfig{file=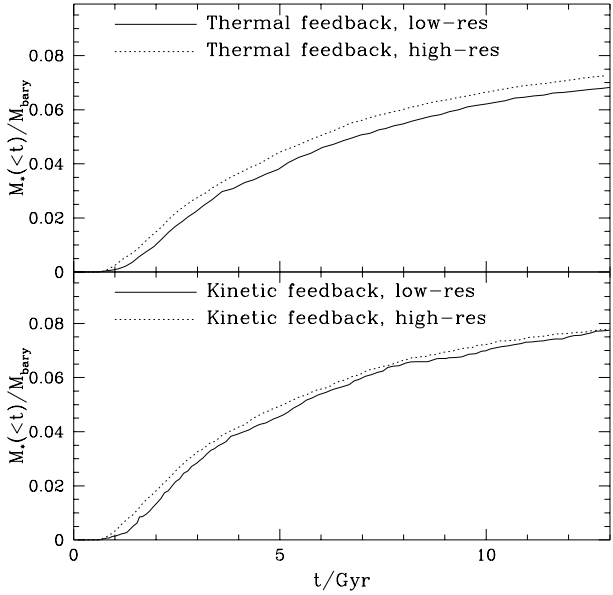,height=8.5cm}}
\caption{Cumulative fraction of baryons in stars against time,
in Gyr, for simulations with thermal and kinetic feedback, and
different mass resolution.}
\label{fig:sf48}
\end{figure}

\begin{figure}
\centering
\centerline{\psfig{file=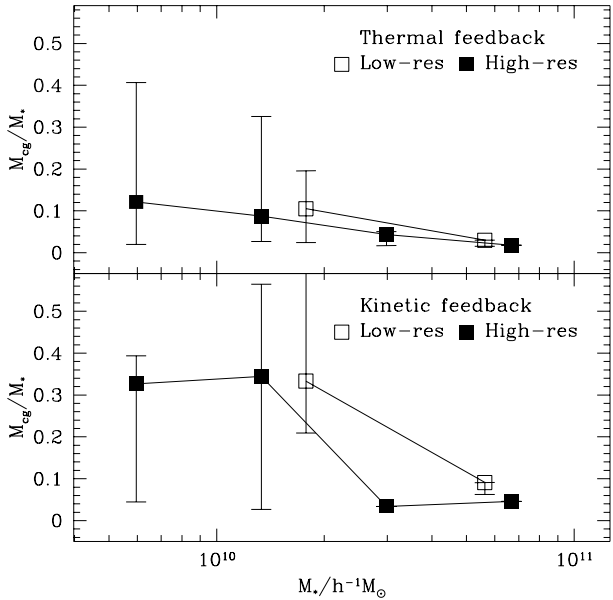,height=8.5cm}}
\caption{Median ratio of cold gas to stars against stellar mass
for galaxies with non-zero gas content and 32 or more particles,
in feedback simulations with varying mass resolution. Error bars 
represent 10 and 90 percentile points in each bin.}
\label{fig:cgf48}
\end{figure}

The main differences in the simulations with low and high resolution
occur at early times and in small haloes. For example, the star
formation histories (Fig.~\ref{fig:sf48}) are quite similar except 
within the first gigayear, when smaller objects are resolved in the high 
resolution simulation.  Increasing the resolution serves 
to extend the range of resolved galaxies to smaller masses, but has 
relatively small effects on more massive galaxies. For example, in
Fig.~\ref{fig:cgf48}, we plot the median ratio of cold gas to stars against
stellar mass for galaxies with non-zero gas content and 32 or more particles 
(error bars represent 10 and 90 percentile points within each bin, of width 
$\Delta \log M_*=0.5$ for low resolution simulations and 0.35 for high
resolution simulations). 
The cold gas fractions in the high resolution simulations are consistent
with the low resolution simulations for galaxies resolved in both.

\begin{figure}
\centering
\centerline{\psfig{file=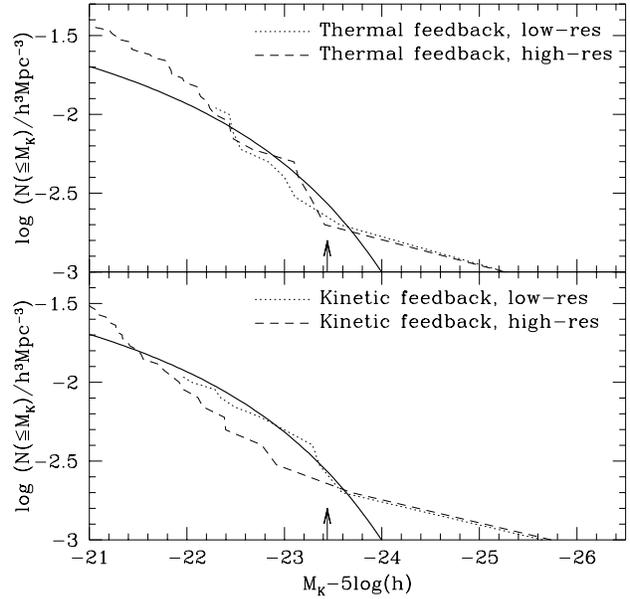,height=8.5cm}}
\caption{$K$--band galaxy luminosity function for the simulations
with thermal and kinetic feedback and varying mass resolution.
Also plotted is the best--fit Schechter function to the
result from the 2dF and 2-Mass surveys (Cole \etal 2001). The value of 
$M_{K}^*$ is illustrated with a vertical arrow.}
\label{fig:lfk48}
\end{figure}

The extension of the resolved galaxy population to smaller masses as
the resolution is increased is clearly illustrated by the $K$-band
galaxy luminosity functions, plotted in Fig.~\ref{fig:lfk48}. In the
thermal feedback case, the luminosity function in the higher
resolution simulation matches well on to the low resolution luminosity
function. In the kinetic feedback case, there is some mismatch around
$M_K^*$, but the number of galaxies is so small here that these differences
are not significant. We conclude that, overall, the increase in the 
resolution of the simulation has little effect on the population of massive
galaxies. 
%{\bf *** SCOTT: Can the galaxies be individually matched in the
%low and high resolution simulations? If so, how do their masses compare?} 

\begin{figure}
\centering
\centerline{\psfig{file=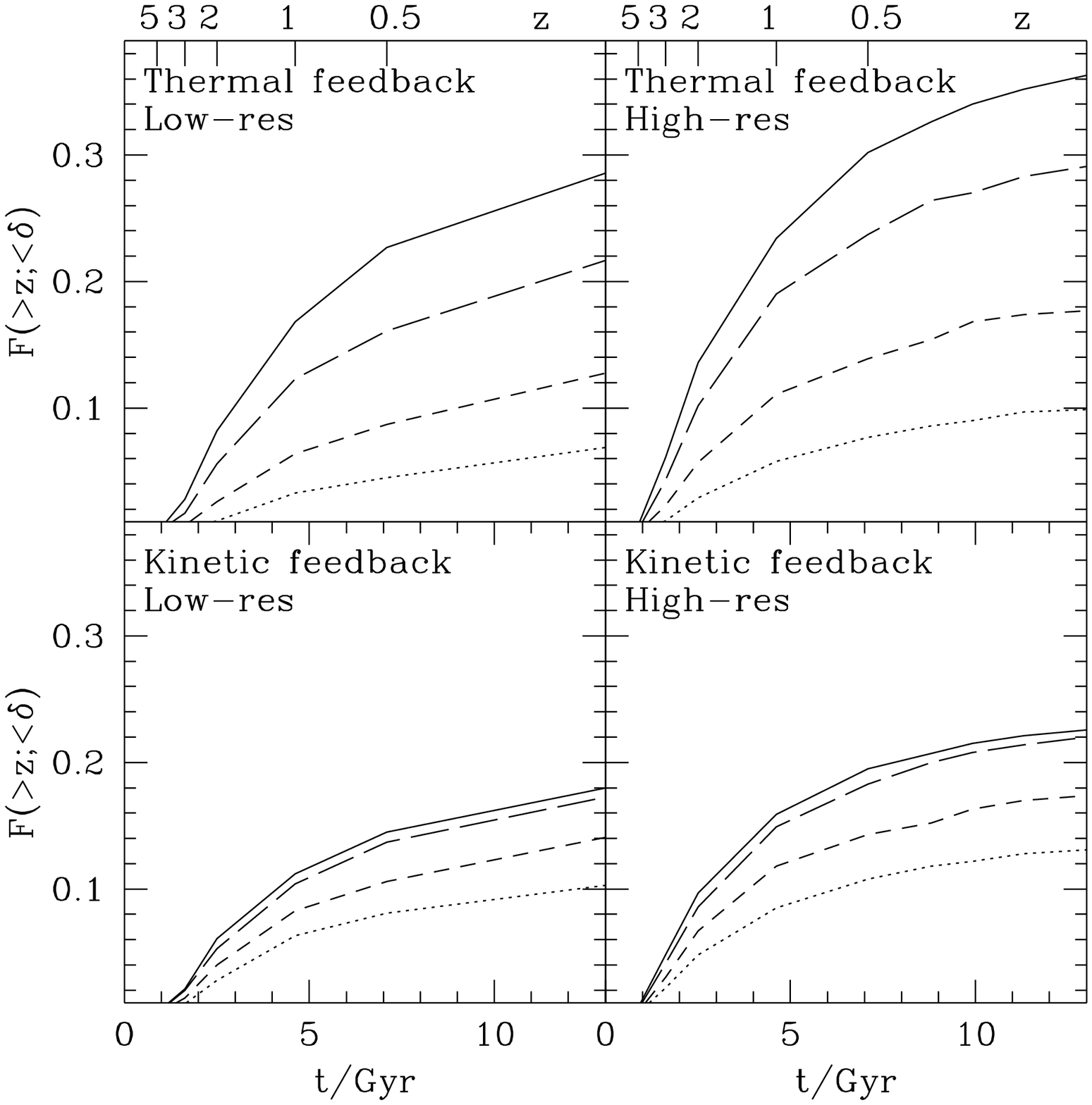,height=8.5cm}}
\caption{Fraction of baryons supplied with supernova energy,
at a given time, for thermal and kinetic feedback simulations with
varying resolution (solid lines). Also plotted are subsets with
$\delta<500$ (long-dashed lines), $\delta<50$ (short-dashed lines) and
$\delta<10$ (dotted lines). Redshift intervals are marked along the 
top of the figure.}
\label{fig:deltaz48}
\end{figure}

\begin{figure*}
\centering
\centerline{\psfig{file=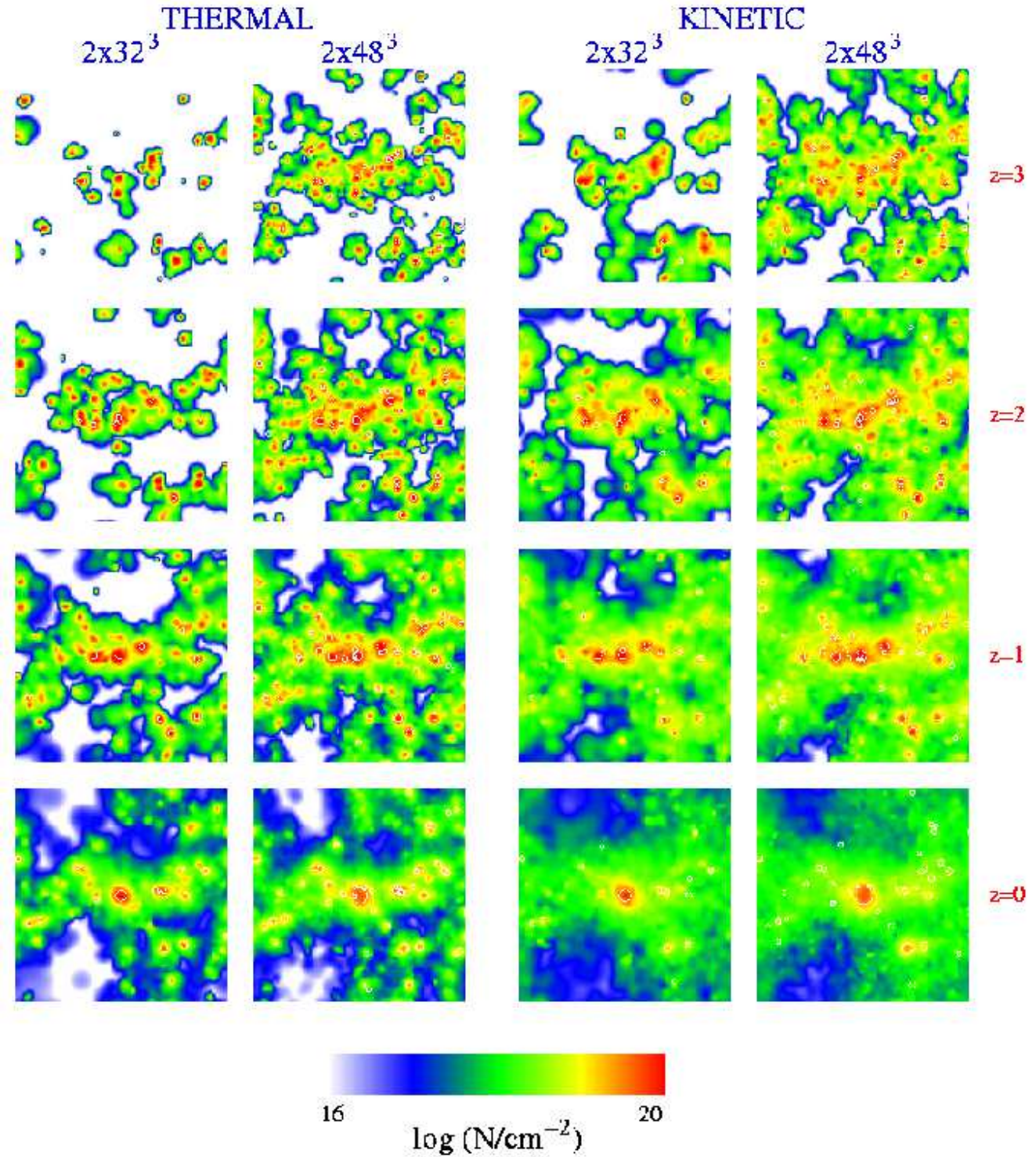,height=20cm}}
\caption{Column density of reheated hydrogen gas in SCDM simulations, with
the positions of the galaxies superimposed. From top to bottom, the rows
are for redshifts 3,2,1 \& 0 respectively. From left to right, the columns
are for thermal feedback in the $32^3$ \& $48^3$ simulations, followed by
kinetic feedback for $32^3$ and $48^3$ simulations. Radii of the circles
are scaled proportional to $M_{\rm gal}^{1/3}$.}
\label{fig:column}
\end{figure*}

Resolution affects are more noticeable in the amount and distribution of
reheated gas. Fig.~\ref{fig:deltaz48} shows the fraction of gas at
different epochs lying in regions of different overdensity (\cf
Fig~\ref{fig:deltaz}).  Increasing the resolution results in an increase in
these fractions at all redshifts. The hydrogen column density distribution
of reheated gas is illustrated in Fig.\ref{fig:column} for the four
simulations, at $z=3,2,1$ \& 0. Column densities were calculated by
smoothing the density of local particles onto a $64\times 64$ grid, using
the SPH kernel with width set by the particles' smoothing lengths.
Overlaid as circles (with radii proportional to $M_{\rm gal}^{1/3}$) are
the positions of galaxies with 32 or more star particles.  At $z=3$ the gas
in the low resolution simulations is confined to the few regions where the
first galaxies have formed (most of these are unresolved.) In the
simulations with higher resolution the gas has a much larger covering
factor because many more galaxies are resolved and so more of the volume is
occupied by galaxies. The range of column densities is high in these
regions (reaching $\sim 10^{20} {\rm cm^{-2}}$).  At lower redshifts, the
distribution of gas becomes much more skewed towards regions of higher
overdensity, as the galaxies and haloes merge into larger units. By $z=0$,
the variation of density contrast in the gas is prominent, although the
distribution is smoother in kinetic than in thermal models. As was inferred
from Fig.~\ref{fig:deltaz}, kinetic feedback is able to transport gas into
regions of lower density faster than thermal feedback, even though both
cause a similar reduction of star formation by $z=0$.

%{\bf *** SCOTT: The caption to Fig 15 is missing.}

%{\bf *** SCOTT: The value of lweight is not consistent from plot to
%plot. For example, Fig 10 has too small a value; Figs 8 and 9 are OK.} 

\section{Conclusions}
\label{sec:sf_conclude}

This paper has surveyed various phenomenological models for
incorporating the effects of star formation and associated energy
feedback (from supernovae) within cosmological simulations of galaxy
formation, using the SPH code {\sc hydra}).  Our main conclusions are
as follows.

(i) Most prescriptions for turning cold gas into stars used in
previous SPH simulations give similar results when adjusted to give
the same present day stellar density, $\Omega_*$. The mass in stars
formed as a function of time is primarily determined by the gas
density at which star formation is assumed to proceed efficiently. 
Thus, adopting
a global density threshold (either in comoving or in physical
coordinates), a Schmidt law, or a more complicated prescription
involving, for example, the Jeans mass, makes little difference to the
global properties of the galaxy population that forms.  Given the
limited resolution of current simulations (in which the internal
properties of galaxies are unresolved), it is simplest to adopt a
global density threshold.

(ii) Incorporating star formation within a cosmological simulation
makes no significant difference to the rate at which gas cools. The
amount of gas cooled by $z=0$ only varies up to 1 per cent in
simulations with and without star formation and the present day
baryonic (gas+stars) mass function of galaxies also remains
unchanged. However, when cold, dense gas particles are replaced by
collisionless particles, a saving of around 50 per cent of the CPU
time is achieved from reducing the (computationally-expensive) work of
calculating hydrodynamical forces.

(iii) Simulations without feedback predict cooled gas fractions of
around 35 per cent. Not only are these values resolution-dependent,
but they exceed observational measurements by around a factor of 5.
The problem of ``overcooling'' is further highlighted in the $K$--band
galaxy luminosity function, which requires the majority of cooled gas
to be ``hidden'' in low-mass non-visible stars (such as brown-dwarfs),
in order to provide a reasonable match to the observed luminosity
function.

(iv) Injecting supernova energy to surrounding gas particles, as
thermal or kinetic energy, can reduce the star formation rate
appreciably.  However, thermal models can only achieve this if the
reheated gas is prevented from cooling for a considerable fraction of
a Hubble time.  For maximal thermal feedback (\ie all available energy
is injected into the gas), the results are less sensitive to the
choice of delay timescale than for models with a lower
efficiency. Average fractions of cold gas in galaxies are unaffected
by the inclusion of feedback.

(v) A significant fraction of the gas reheated by supernovae is able
to escape galaxy haloes and move into the low density regions
($\delta<10$) responsible for the Ly-$\alpha$ forest. If this gas is
viewed as a tracer of reprocessed material, this mechanism efficiently 
transports metals into low-density regions.

(vi) The efficiency with which supernovae expel gas from haloes depends
on the circular velocity of the halo ( $100-300 \kms$ in these simulations) 
- more gas is expelled from smaller haloes, particularly for models capable 
of strong feedback.  Larger simulations are required to establish robustly 
the form of this relation.

(vii) Increasing the mass resolution by about a factor of $\sim 3$
increases the total amount of resolved galactic material primarily
because smaller galaxies are then resolved.  Consequently, the
$K$--band galaxy luminosity function is very similar over the range of
magnitudes resolved in our high and low resolution simulations.  The
ratio of cold gas to stars is also essentially unaffected, as is
distribution of gas perturbed by supernovae, although increasing the
resolution increases the amount of this material, due to the higher
number of resolved galaxies.  Hence, until numerical convergence is
established, the simulations underpredict the amount of this material.

To fully test the effectiveness of the feedback schemes that
we have explored in this paper will require larger simulations to be
performed in statistically significant volumes. This will allow
definite predictions to be made for \eg the galaxy luminosity function
and correlation function, which can readily be compared with
observations.

\section*{Acknowledgments}
We would like to thank Andrew Benson, George Efstathiou, Cesario Lia, 
Rob Thacker, Peter Thomas, Tom Theuns, David Weinberg and Simon White for 
helpful discussions. We also thank the referee for constructive comments
that have improved the clarity of this paper.
STK acknowledges PPARC for financial support.
The work presented in this paper was carried out as part of the programme
of the Virgo Supercomputing Consortium 
({\tt http://star--www.dur.ac.uk/$\sim$frazerp/virgo/}). 
This work was supported by the EC TMR network for ``Galaxy 
formation and evolution''.


\begin{thebibliography}{}
\bibitem[\protect\citename{Balogh \etal }2001]{Balogh2001} Balogh M.~L., Pearce F.~R., Bower R.~G., Kay S.~T., 2001, \mnras, in press (astro-ph/0104041)
\bibitem[\protect\citename{Baugh \etal }2001]{Baugh01} Baugh C.~M., Benson A.~J., Cole S., Frenk C.~S., Lacey C.~G., 2001, proceedings of ``QSO hosts and their environments'', IAA, Granada (astro-ph/0103156)
\bibitem[\protect\citename{Bardeen et al. }1986]{BBKS} Bardeen J.~M., Bond J.~R., Kaiser N., Szalay A.~S., 1986, \apj, 304, 15
\bibitem[\protect\citename{Benson \etal }2001]{Benson01} Benson A.~J., Frenk C.~S., Baugh C.~M., Cole S., Lacey C.~G., 2001, \mnras, submitted (astro-ph/0103092)
\bibitem[\protect\citename{Benz }1990]{BENZ} Benz W., 1990, in Buchler J.~R., ed., Numerical Modelling of Stellar Pulsations: Problems and Prospects. Kluwer, Dordrecht, p. 269
\bibitem[\protect\citename{Bond \& Efstathiou }1984]{BE} Bond J.~R., Efstathiou G., 1984, \apj, 285, L45
\bibitem[\protect\citename{{Bruzual} \& {Charlot} }1993]{BC93} {Bruzual A.} G.,  {Charlot} S., 1993, \apj, 405, 538
\bibitem[\protect\citename{Buonomo \etal }2000]{BUONOMO99} Buonomo F., Carraro G., Chiosi C., Lia C., 2000, \mnras, 312, 371
\bibitem[\protect\citename{Carraro \etal }1998]{Carraro98} {Carraro} G., {Lia} C.,  {Chiosi} C., 1998, \mnras, 297, 1021
\bibitem[\protect\citename{Cen \& Ostriker }1992]{CO92} Cen R., Ostriker J., 1992, \apj, 393, 22
\bibitem[\protect\citename{Cen \& Ostriker }1999]{CO99} Cen R., Ostriker J., 1999, \apj, 538, 83
\bibitem[\protect\citename{Cole }1991]{Cole91}Cole S., 1991, \apj, 367, 45
\bibitem[\protect\citename{{Cole} et~al.}{1994}]{Cole94}{Cole} S., {Aragon-Salamanca} A., {Frenk} C.~S., {Navarro} J.~F.,  {Zepf} S.~E., 1994, \mnras, 271, 781
\bibitem[\protect\citename{Cole \etal}{2000}]{Cole2000}Cole S., Lacey C.~G., Baugh C.~M., Frenk C.~S., 2000, \mnras, 319, 168
\bibitem[\protect\citename{Cole \etal}{2001}]{Cole2001}Cole S., Norberg P., Baugh C., Frenk C., \etal (The 2dFGRS Team), 2001, \mnras, in press (astro-ph/0012429)
\bibitem[\protect\citename{Couchman }1991]{COUCH} Couchman H.~M.~P., 1991, \apj, 368, L23
\bibitem[\protect\citename{Couchman, Thomas \& Pearce }1995]{CTP} Couchman H.~M.~P., Thomas P.~A., Pearce F.~R., 1995, \apj, 452, 797
\bibitem[\protect\citename{Croft \etal }2000]{Croft2000} Croft R.~A.~C., Di Matteo T.,
Dav\'{e} R., Hernquist L., Katz N., Fardal M.~A., Weinberg D.~H., 2000, \apj, submitted (astro-ph/0010345)
\bibitem[\protect\citename{Davis \etal }1985]{DEFW} Davis M., Efstathiou G., Frenk C.~S., White S.~D.~M., 1985, \apj, 292, 371
\bibitem[\protect\citename{{Dekel} \& {Silk} }{1986}]{Dekel&Silk86}{Dekel} A.,  {Silk} J., 1986, \apj, 303, 39
\bibitem[\protect\citename{Efstathiou }{1992}]{Efstathiou1992} Efstathiou G., 1992, \mnras, 256, 43
\bibitem[\protect\citename{Efstathiou }{2000}]{Efstathiou2000} Efstathiou G., 2000, \mnras, 317, 697
\bibitem[\protect\citename{Eke, Cole \& Frenk }1996]{ECF} Eke V.~R., Cole S.~M., Frenk C.~S., 1996, \mnras, 282, 263
\bibitem[\protect\citename{Eke, Navarro \& Frenk }1998]{ENF98} Eke V.~R., Navarro J.~F., Frenk C.~S., 1998, \apj, 503, 569
\bibitem[\protect\citename{Ellison \etal }1999]{Ellison99} Ellison S.~L., Lewis G.~F., Pettini M., Chaffee F.~H., Irwin M.~J., 1999, \apj, 520, 456
\bibitem[\protect\citename{Evrard, Summers \& Davis }1994]{ESD} Evrard A.~E., Summers F.~J., Davis M., 1994, \apj, 422, 11
\bibitem[\protect\citename{Fardal \etal }{2000}]{Fardal2000}Fardal
M.~A., Katz N., Gardner J.~P., Hernquist L., Weinberg D.~H., Dav\'{e}
R., 2000, \apj, submitted (astro-ph/0007205) 
\bibitem[\protect\citename{Frenk \etal }1996]{FEWS} Frenk C.~S., Evrard A.~E., 
White S.~D.~M., Summers F.~J., 1996, \apj, 472, 460
\bibitem[\protect\citename{{Gerritsen} }{1997}]{G97}Gerritsen
J.~P.~E., 1997, PhD Thesis, Groningen University, the Netherlands 
\bibitem[\protect\citename{Granato \etal }2000]{Granato00} Granato
G.~L., Lacey C.~G., Silva L., Bressan A., Baugh C.~M., C ole S., Frenk
C.~S., 2000, \apj, 542, 710 
\bibitem[\protect\citename{Katz }{1992}]{K92}Katz N., 1992, \apj, 391, 502
\bibitem[\protect\citename{Katz, Hernquist \& Weinberg }1992]{KHW}
Katz N., Hernquist L., Weinberg D.~H., 1992, \apj, 399, L109 
\bibitem[\protect\citename{{Katz}, {Weinberg}, \& {Hernquist}
}{1996}]{KWH96}{Katz} N., {Weinberg} D.~H.,  {Hernquist} L., 1996,
\apjs, 105, 19 
\bibitem[\protect\citename{{Kay} et~al }{2000}]{Kay2000}{Kay} S.~T.,
{Pearce} F.~R., Jenkins A., Frenk C.~S., White S.~D.~M., Thomas P.~A.,
Couchman H.~M.~P., 2000, \mnras, 316, 374 (K2000) 
\bibitem[\protect\citename{{Kauffmann} et~al. }{1994}]{GK94}{Kauffmann}
G., {Guiderdoni} B.,  {White} S.~D.~M., 1994, \mnras, 267, 981 
\bibitem[Kauffmann, Nusser \& Steinmetz 1997]{kns}Kauffmann~G.,
Nusser~A., Steinmetz~M., 1997, MNRAS, 286, 795 
\bibitem[Kauffmann \etal 1999a]{kauff99a}Kauffmann G., Colberg
J.~M., Diaferio A., White S.~D.~M., 1999a, MNRAS, 303, 188 
\bibitem[Kauffmann \etal 1999b]{kauff99b}Kauffmann G., Colberg
J.~M., Diaferio A., White S.~D.~M., 1999b, MNRAS, 307, 529 
\bibitem[\protect\citename{Kravtsov, Klypin \& Khokhlov }1998]{KKK98}
Kravtsov A.~V., Klypin A.~A., Khokhlov A.~M., 1998, American
Astronomical Society Meeting, 193, 3908 
\bibitem[\protect\citename{{Kennicutt}
}{1983}]{Kennicutt83}{Kennicutt} R.~C., 1983, \apj, 272, 54 
\bibitem[\protect\citename{{Kennicutt} }{1998}]{KEN98}{Kennicutt}
R.~C., 1998, \apj, 498, 541 
\bibitem[\protect\citename{Lacey \& Cole }1994]{LC94} Lacey C., Cole
S., 1994, \mnras, 271, 676 
%\bibitem[\protect\citename{{Loveday} }{2000}]{Loveday2000}
%{Loveday} J., 2000, \mnras, 312, 557
\bibitem[\protect\citename{Mac Low \& Ferrara }{1999}]{Maclow&Ferrara} Mac Low M.~M., Ferarra A., 1999, \apj, 513, 142
\bibitem[\protect\citename{Madau, Ferrara \& Rees }{2000}]{Madau2000} Madau P., Ferrara A., Rees M.~J., 2000, \apj, submitted (astro-ph/0010158)
\bibitem[\protect\citename{Martin }{1999}]{Martin1999} Martin C.~L., 1999, \apj, 513, 156
\bibitem[\protect\citename{Mihos \& Hernquist }{1994}]{Mihos&Hernquist}Mihos J.~C., Hernquist L., 1994, \apj, 437, 611
\bibitem[\protect\citename{{Miller} \& {Scalo} }{1979}]{Miller&Scalo79}{Miller} G.~E.,  {Scalo} J.~M., 1979, \apjs, 41, 513
\bibitem[\protect\citename{Monaghan }1992]{MON} Monaghan J.~J., 1992, \ARAA, 30, 543
\bibitem[\protect\citename{Navarro \& White }{1993}]{NW93}Navarro J.~F.,  White S.~D.~M., 1993, \mnras, 265, 271
\bibitem[\protect\citename{Navarro \& Steinmetz }{1997}]{NS97}Navarro J.~F., Steinmetz M., 1997, \apj, 478, 13
\bibitem[\protect\citename{Pearce }{1998}]{PSF}Pearce F.~R., 1998, astro-ph/9803133
\bibitem[\protect\citename{{Pearce} et~al. }{1999}]{Pearce99}{Pearce}
F.~R. et~al., 1999, \apj, 521, L99 
\bibitem[\protect\citename{{Pearce} et~al. }{2000}]{Pearce2000} Pearce
F.~R., Thomas P.~A., Couchman H.~M.~P.,  Edge A.~C., 2000, \mnras,
317, 1029 
\bibitem[\protect\citename{Pettini \etal }{2001}]{Pettini2001} Pettini M., Shapley A.~E., Steidel C., Cuby J.-G., Dickinson M., Moorwood A.~F.~M., Adelberger K.~L., Giavalisco M., 2001, \apj, 554
\bibitem[\protect\citename{Ritchie \& Thomas }{2001}]{Ritchie&Thomas}
Ritchie B.~W., Thomas P.~A., 2001, \mnras, 323, 743 
\bibitem[\protect\citename{Quinn, Katz \& Efstathiou
}{1996}]{QKE}Quinn T.~R., Katz N., Efstathiou G., 1996, \mnras, 278,
L49 
\bibitem[\protect\citename{Schaye \etal }2000]{Schaye2000} Schaye J.,
Rauch M., Sargent W.~L.~W., Kim T., 2000, \apj, 541, L1 
\bibitem[\protect\citename{{Schmidt} }{1959}]{schmidt}{Schmidt} M.,
1959, \apj, 129, 243 
\bibitem[\protect\citename{Silk \& Rees }{1998}]{Silk&Rees98} Silk J.,
Rees M.~J., 1998, \aap, 331L, 1
\bibitem[Somerville \etal 2000]{somerville00}Somerville~R.~S.,
Lemson~G., Sigad~Y., 
Dekel~A., Kauffmann~G., White~S.~D.~M., submitted to MNRAS
(astro-ph/9912073)
\bibitem[\protect\citename{Somerville \& Primack }{1999}]{SP98}Somerville R.~S.,  Primack J.~R., 1999, \mnras, 310, 1087
\bibitem[\protect\citename{Springel }{2000}]{Springel2000} Springel V., 2000,
\mnras, 312, 859
\bibitem[\protect\citename{Steinmetz \& M\"uller }{1995}]{SM95}Steinmetz M.,  M\"uller E., 1995, \mnras, 276, 549
\bibitem[\protect\citename{Sugiyama }1995]{Sugi95} Sugiyama N., 1995, \apjs, 100, 281
\bibitem[\protect\citename{{Summers} }{1993}]{Summers_phd}{Summers} F.~J., 1993, Ph.D. thesis, University of California 
\bibitem[\protect\citename{Sutherland \& Dopita }{1993}]{SUTDOP}Sutherland R.~S.,  Dopita M.~A., 1993, \apjs, 88, 253
\bibitem[\protect\citename{Thacker \& Couchman }{2000}]{Thacker&Couchman2000}Thacker R.~J., Couchman H.~M.~P., 2000, \apj, 545, 728
\bibitem[\protect\citename{Thomas \& Couchman }1992]{THOMCOUCH} Thomas P.~A., Couchman H.~M.~P., 1992, \mnras, 257, 11 
\bibitem[\protect\citename{Theuns \etal }1998]{Theuns98} Theuns T., Leonard A., Efstathiou G., Pearce F.~R., Thomas P.~A., 1998, \mnras, 301, 478
\bibitem[\protect\citename{Theuns, Mo \& Schaye }2001]{Theuns2000} Theuns T., Mo H.~J., Schaye J., 2001, \mnras, 321, 450
\bibitem[\protect\citename{Thoul \& Weinberg }1996]{Thoul&Weinberg96} Thoul A.~A., Weinberg D.~H., 1996, \apj, 465, 608
\bibitem[\protect\citename{{Tissera}, {Lambas}, \& {Abadi} }{1997}]{TISSERA}{Tissera} P.~B., {Lambas} D.~G.,  {Abadi} M.~G., 1997, \mnras, 286, 384
\bibitem[\protect\citename{Vianna \& Liddle }1996]{VL} Viana P.~T.~P., Liddle A.~R., 1996, \mnras, 281, 323
\bibitem[\protect\citename{Weinberg, Hernquist \& Katz} 1997]{WHK97} Weinberg D.~H., Hernquist L.,Katz N., 1997, \apj, 477, 8
\bibitem[\protect\citename{White \& Frenk }{1991}]{WF} White S.~D.~M.,  Frenk C.~S., 1991, \apj, 379, 52
\bibitem[\protect\citename{White \& Rees }{1978}]{WR} White S.~D.~M.,  Rees M.~J., 1978, \mnras, 183, 341
\bibitem[\protect\citename{Yepes \etal }{1997}]{YKKK} Yepes G., Kates R., Khokhlov A., Klypin A., 1997, \mnras, 284, 235 
\end{thebibliography}
\end{document}